\documentclass[paper=a4, 12pt]{article}
\usepackage{floatrow}
\newfloatcommand{capbtabbox}{table}[][\FBwidth]

\usepackage{algorithm}
\usepackage[noend]{algpseudocode}

\usepackage{epsfig,amsmath,enumerate,amsfonts,xspace}
\usepackage{latexsym}
\usepackage{amsmath}
\usepackage[parfill]{parskip} 
\usepackage[top=2.5cm, bottom=2.5cm, left=2.5cm, right=2.5cm]{geometry}
\usepackage{amssymb,amsthm,amsbsy,amsfonts}
\usepackage{hyperref}
	\hypersetup{
    colorlinks = false
}
\usepackage{natbib}
\usepackage{multirow, booktabs}

\usepackage{graphicx}
\usepackage{color}
\usepackage{subfig}
\usepackage{psfrag}
\usepackage{pdflscape}
 \usepackage{float}

\newcommand{\mat}[1]{\ve #1}
\newcommand{\abs}[1]{\left | #1 \right |}

\newcommand{\tr}{\mbox{tr}} 
\newcommand{\T}{T} 

\newcommand{\cov}{\mbox{Cov}}
\newcommand{\corr}{\mbox{corr}}
\newcommand{\E}{\mbox{E}}
\newcommand{\ve}[1]{\mbox{\boldmath ${#1}$}}
\newcommand{\vesub}[2]{\mbox{{\boldmath ${#1}$}$_{#2}$}}

\newcommand{\CN}{\mathcal N}

\newcommand{\R}{\mbox{${\cal R}$}}
\newcommand{\GP}{\mbox{${\cal GP}$}}

\begin{document}

\title{Bayesian analysis of nonlinear structured latent factor  models using a Gaussian Process Prior}   
%
%
\author{Yimang Zhang\textsuperscript{1}, Xiaorui Wang\textsuperscript{1} and Jian Qing Shi\textsuperscript{1,2} \footnote{Corresponding author, email: shijq@sustech.edu.cn.} \\[3mm]
	\small{\textsuperscript{1}Department of Statistics and Data Science, 
		Southern University of Science and Technology, China}\\
	\small{\textsuperscript{2}National Center for Applied Mathematics Shenzhen, China}
	}

\maketitle
\begin{abstract} 
 Factor analysis models are widely utilized in social and behavioral sciences, such as psychology, education, and marketing, to measure unobservable latent traits. In this article, we introduce a nonlinear structured latent factor analysis  model which is more flexible to characterize the relationship between manifest variables and latent factors. The confirmatory identifiability of the latent factor is discussed,  ensuring the substantive interpretation of the latent factors.
A Bayesian approach with a Gaussian process prior
is proposed to estimate the unknown nonlinear function and the unknown parameters.
Asymptotic results are established, including structural identifiability of the latent factors, consistency of the estimates of the unknown  parameters and the unknown nonlinear function. Simulation studies and a real data analysis are conducted to investigate the performance of the proposed method. Simulation studies show our proposed method performs well in handling nonlinear model and successfully identifies the latent factors. Our analysis incorporates oil flow data, allowing us to uncover the underlying structure of  latent nonlinear patterns.
\end{abstract}

%
%
%
%
%
%
%

\newpage

\section{Introduction}

Factor analysis \citep{McCabe:1984} is widely employed in fields such as psychology, social sciences, and market research, serving as a useful tool for modeling  common dependence among multivariate data. 
Identifiability is a crucial property of latent factor models, essential for ensuring a substantive interpretation of latent factors. When a model is identifiable, certain latent factors can be uniquely extracted from observed data, facilitating a rich understanding of the relationships between observed variables and latent factors. It is well known that a factor model is typically non-identifiable due to it's invariance with respect to orthogonal transformations. Some existing methods achieve identifiability by imposing constraints on the factor loadings. Such as, 
Varimax \citep{rohe2020vintage}, Promax and Oblimin  \citep{abdi2003factor} methods achieve identifiability by maximizing the simplicity or sparsity of factor loadings; Several approaches \citep{bartholomew2011latent} facilitate the identification of remaining parameters by setting a specific factor loading to zero or fixing the variance of a particular latent factor.

In recent years, there has been widespread interest in investigating the identifiability of factor models based on the design information between observed variables and latent factors, which is commonly applied in the field of confirmatory factor analysis \citep{harrington2009confirmatory}. Such approaches involve pre-specifying design information and then translating it into a zero constraint on a loading parameter of the model. Therefore, it is also referred to as structural latent factor model. \cite{chen2020structured}
studied how the design information affects the identifiability and the
estimation of a generalized structured latent factor model, incorporating linear latent factor models. \cite{leeb2021note}  gave an elementary proof
showing that the asymptotic identifiability of latent factors proposed by \cite{chen2020structured} holds non-asymptotically, and provided the identifiability of the loading matrix.
\cite{papastamoulis2022identifiability} studied the identifiability of Bayesian factor analytic models. 

The existing methods are based on the assumption of linear or generalized linear relationship between manifest variables and latent factors, but in real-world data the relationship is usually unknown. In this paper, we consider a nonlinear structural latent factor model, where the link function is unknown.
Due to the complex distribution associated with the nonlinear factors, statistical analyses of
nonlinear factor analysis models are very difficult. Nonlinear FA models with polynomial relationships were first explored
by \cite{mcdonald1962general}. 
\cite{yalcin2001nonlinear} 
proposed a more general nonlinear factor analysis model. However, the assumption made in this model is considered too strong because it explicitly specifies the nonlinear relationships and assumes that the factor represents the “true value” of certain observed variables, thereby eliminating any factor indeterminacy. The study only considers a single factor model, which limits its generality, and the parametrization of measurement errors in the variables does not pose any identifiability issues.
\cite{paisley2009nonparametric}  propose a non-parametric extension to the linear
factor analysis  using a beta process
prior on the loading matrix to achieve its sparseness, which has been introduced for dictionary learning of the sparse representation in speech enhancement \citep{li2016speech}. Other process can also be used as a prior like  Gamma process dynamic poisson factor
analysis model \citep{acharya2015nonparametric} and  negative binomial factor analysis model \citep{zhou2018nonparametric}.

Gaussian Process latent variable models (GPLVM) \citep{Lawrence:2005} use the idea of Gaussian Process regression(GPR) model \citep{Shi:2011} to do non-parametric and non-linear dimension reduction. Gaussian Process Dynamical Models (GPDM) \citep{wang2005gaussian}
assumes that different samples are  correlated by time, in this model, the unknown link function is also assumed a  GP  priori.  \cite{damianou2011variational} gives  another GP prior in the latent space. However, the computation of the inverse of $N \times N$ covariance matrix is very heavy, where $N$ is the number of recording time points and \cite{damianou2011variational} uses variational approximations for GPDM. The limit is that the approximation of Variational Bayesian cannot reach consistency of the estimates.  Although both GPLVM and GPDM 
are capable of making excellent predictions, they do not take into account identifiability of latent variables. Non-identifiability of latent factors limits their interpretability, and thus limits their applications for some real-word problems. We will discuss this problem with an unknown link function.   


In this paper, we discuss the identifiability and estimation of a nonlinear structured factor analysis model. The effectiveness of the proposed method is validated through asymptotic properties and empirical studies. Comparing with exsiting works, our paper has following contributions. First, We propose a nonlinear structural factor model with an unknown linking function. In comparison to conventional structural factor models, it offers high flexibility in practical applications. Furthermore, in contrast to single- and multiple-index models, we extend nonlinear functions to the domain of factor models. Second, we apply nonparametric Bayesian methods to latent factor models, specifically employing GPR approach to estimate unknown functions. Unlike GPLVM, which does not provide a clear substantive  interpretation of latent factors, we establish identifiability for latent factors.

The rest of the article is organized as follows. In Section \ref{main results}, we define  nonlinear structured latent factor models, investigate the  identifiability and provide  two estimators of the 
unknown parameters, latent factors and unknown link functions. In Section \ref{identifiability}, we discuss the identifiability of the latent factors, the consistency of the estimates of unknown parameters and posterior consistency of the estimates of the unknown link functions. In
Section \ref{Numerical}, numerical results are presented that contain  simulation studies and an application
to a  oil dataset. Finally, concluding remarks and further studies 
are made in Section \ref{concluding}. All technical proofs and additional simulated examples are provided in Appendix.

\section{Model description: basic setup and implementation}\label{main results}
\subsection{Nonlinear structured latent factor  model} \label{Themodel}

A structured latent factor model refers to a statistical model that represents the relationships among manifest (observed) variables through low-dimensional latent (unobserved) factors. The term “structured” indicates that the model incorporates a specific pattern or constraints on the relationships among the latent factors and observed variables which is crucial for ensuring the identifiability of the latent factors. Conventionally, such a structure is established by imposing zero constraints on factor loadings, indicating that each observed variable depends on certain latent factors. Consequently, these latent factors can be uniquely extracted from the data, ensuring both identifiability and substantive interpretability. 
\cite{chen2020structured} investigated the identifiability and estimation of structured factor models with given generalized link functions, including linear and logistic models as  special cases.
However, in practice, the relationship between observed variables and latent factors may be unknown. In this paper, we consider a nonlinear structured latent factor  model (NSLFA). 

Denote $N$ as the number of individuals (e.g., $N$ test-takers) and $J$ as the number of
manifest variables (e.g., $J$ test items). Let $Y_{ij}$ be  a random variable denoting the $i$th individual’s value on the $j$th manifest variable.  Suppose that each individual $i$ is associated with a $K$-dimensional latent factor, denoted as $\boldsymbol{x}_i=\left(x_{i 1}, \ldots, x_{i K}\right)^{\top}$ and each manifest variable $j$ is associated with $K$ parameters $\ve{a}_j=\left(a_{j 1}, \ldots, a_{j K}\right)^{\top}$. The model can be written as
\begin{eqnarray}
Y_{ij}  =f_j(\ve{a}_j^{\top} \boldsymbol{x}_i)+\varepsilon_{ij}, \ \ \varepsilon_{ij} \sim \CN(0, \sigma^2), \  \ \text{for } j=1,\ldots, J,  \label{GPSIMFA}
\end{eqnarray}
where $f_1, \cdots, f_J$ are unknown link function and $\varepsilon_{ij}$'s are i.i.d. random errors. 
When $f_j(t)=t$, i.e., identity link function for all $j=1, \ldots, J$, the model degenerates into a linear factor analysis model.
When $f_j(t)$'s are the inverse of  known link function, like  logistic or log link function, the model degenerates into a generalized linear factor analysis model \citep{chen2020structured}.

Model (\ref{GPSIMFA}) extends the single- or multiple-index model to latent factor analysis. In single index models, there are various methods to estimate unknown link functions. The frequentist typically employs methods such as kernel, local linear, B-spline, while in the Bayesian framework, the GPR method is commonly used. 
In this paper, we adapt a Bayesian Non-parametric method  with a Gaussian process prior or  T-processs prior \citep{choi2011gaussian,wang2021general} to estimate the unknown function.

As mentioned before we will study the identifiability of
the structured nonlinear factor analysis model. To this end,
design information
is incorporated as zero constraints on loading coefficients $\ve{a}_j\ \text{for } j=1,\ldots, J  $. Design information is recorded by a matrix $Q=(q_{jk})_{J \times K}$, where each entry takes value 0 or 1. "0" indicates that the $j$-th manifest variable is not directly associated with the $k$-th latent factor. 
Under the confirmatory manner of latent factor models, design information on the relationship between manifest variables and latent factors is previously known.  A design matrix $Q$ can be defined to incorporate domain knowledge into the statistical analysis by imposing zero constraints on factor loadings, i.e., the loading parameters $a_{j k}$ is constrained to zero if $q_{j k}$ is zero and otherwise if $q_{j k}$ is 1. The constraints induced by the design matrix play an important role in the identifiability and the interpretation of the latent factors. We now  discuss the identifiability and how to estimate the unknown quantities in the following subsections. Before proceeding further, let's introduce some notations. For positive integers $K, N$ and $J$, let $\ve x_1, \ldots, \ve x_N$ be vectors in $\mathbb{R}^K$ and let $\ve a_1, \ldots, \ve a_J$ be vectors in $\mathbb{R}^K$; and define the matrices $\ve X^{\top}=\left[\ve x_1, \ldots, \ve x_N\right] \in \mathbb{R}^{K \times N}$ and $\ve A^{\top}=\left[\ve a_1, \ldots, \ve a_J\right] \in \mathbb{R}^{K \times J}$. Let $\mathbb{Z}_{+}$ denote the set of positive integers; $\mathbb{R}^K$ and $\mathbb{R}^{\mathbb{Z}_{+}}$ represent vectors with $K$ and countably infinite real number components, respectively; $\mathbb{R}^{K \times N}$, $\mathbb{R}^{K \times J}$ and $\mathbb{R}^{\mathbb{Z}_{+}\times K}$ respectively represent matrices of size $K \times N$, $K \times J$ and countably infinite rows by $K$ columns.

\subsection{Structural Identifiability}

To ensure the substantive interpretation of latent factors, identifiability plays a pivotal role in structured latent factor model. Under Model (\ref{GPSIMFA}), it is well known that the latent factor is not identifiable due to rotational indeterminacy. To this end, we first formalize the definition of structural identifiability. Here we consider that the number of individuals $N$ grows to infinity and the number of manifest variables $J$ is fixed, then the identifiability of the $k$-th latent factor $\ve X_{[k]}=(x_{1k},x_{2k},\ldots)^{\top}\in\mathbb{R}^{\mathbb{Z}_{+}}$ is equivalent to the identification of the direction of an infinite dimensional vector. Define the following  as
$$
\sin _{+} \angle(\ve{u}, \ve{v})=\limsup _{N \rightarrow \infty} \sin \angle\left(\ve{u}_{[1: N]}, \ve{v}_{[1: N]}\right) \text {, }
$$
to quantify the angle between  two vectors $\ve{u}$ and $\ve{v}$ where $\ve{u}=$ $\left(u_1, u_2, \ldots\right)^{\top}, \ve{v}=\left(v_1, v_2, \ldots\right)^{\top} \in \mathbb{R}^{\mathbb{Z}_{+}}$ are two vectors with countably  infinite components. When $\sin _{+} \angle(\ve{u}, \ve{v})$ is 0, we say the angle between $\ve{u}$ and $\ve{v}$ is 0.

\textbf{Definition 1} (Structural identifiability of a latent factor). Consider the $k$-th latent factor, where $k \in \Bar{k}= \{1, \ldots, K\}$, and a nonempty parameter space $\mathcal{S} \subset \mathbb{R}^{\mathbb{Z}_{+}\times K} \times \mathbb{R}^{J\times K}$ for $(\ve X, \ve A)$ defined in the Appendix \ref{A1A2}. We say the $k$-th latent factor is structurally identifiable in the parameter space $\mathcal{S}$ if for any $(\ve X, \ve A),\left(\ve X^{\prime}, \ve A^{\prime}\right) \in \mathcal{S}, P_{\ve X, \ve A}=P_{\ve X^{\prime}, \ve A^{\prime}}$ implies $\sin _{+} \angle\left(\ve X_{[k]}, \ve X_{[k]}^{\prime}\right)=0$, where $P_{\ve X, \ve A}$ is the probability distribution of $\left(Y_{i j}, i\in \mathbb{Z}_{+},\ j\in\{1,\ldots, J\}\right)$, given factor scores $\ve X$ and loadings   $\ve A$.

The following Theorem 1 provides a necessary and sufficient condition on the design matrix $Q$ for the
structural identifiability of the $k$-th latent factor.

\textbf{Theorem 1.}  Denote $R_Q(S)=\left\{j: q_{j k}=1, \text { if } k \in S \text { and } q_{j k}=0,\right. 
\text { if } k \notin S, 1 \leq j \leq J\}$, where $S$ is a subset of $\Bar{k}$. Under the Definition 1 and Assumptions A1-A2 given in the Appendix \ref{A1A2}, given the hyper-parameters $\tau, w$, the $k$-th latent factor is structurally identifiable in $\mathcal{S}_Q$ if and only if
\begin{equation}
    \{k\}=\bigcap_{k \in S, \ R_Q(S)\ \text{non-empty}} S,
    \label{Thm1}
\end{equation}
where we define $\bigcap_{k \in S, \ R_Q(S)\ \text{non-empty}} S=\emptyset$ if $R_Q(S)=\emptyset$ for all $S$ that contains $k$.

The detailed setting of the model including the definition of hyper-parameters will be given in the following subsections. The proof of theorem is given in Appendix \ref{pf1}.

\subsection{Estimation of unknown parameters, latent factors and unknown function}
\label{Estimation}
Before presenting the estimates, we give a very brief  introduction to the GPR approach. Both Bayesian non-parametric methods with a GP prior or a T-process prior are flexible and can capture complex patterns, the key distinction lies in the choice of the underlying distribution. A GP prior assumes a Gaussian distribution, while a T-process prior incorporates heavy-tailed distributions like the Student's t-distribution to handle non-Gaussian data and outliers more effectively.

The GPR model is a nonparametric model and has some nice features; see details in \citep{Shi:2011}. Suppose we have a data set
$$
\mathcal{D}=\left\{\left(\begin{array}{c}
Y_1 \\
\boldsymbol{t}_1
\end{array}\right)\left(\begin{array}{l}
Y_2 \\
\boldsymbol{t}_2
\end{array}\right) \ldots\left(\begin{array}{c}
Y_N \\
\boldsymbol{t}_N
\end{array}\right)\right\} .
$$
In general, $\ve t \in \R^d$ is a $d$-dimensional vector but in this paper $t$ is a scalar index value. Here we take the following model  
$$Y=f(\ve t)+\varepsilon, \ \ \varepsilon \sim \CN(0, \sigma^2) $$ 
as an example to illustrate the GPR approach and a Gaussian process is used as a prior of the unknown function $f(\cdot)$ in a functional space, i.e., 
\begin{equation}
    f(\cdot)  \overset{prior}{\mathop{\sim}}\GP \left( 0,k(\cdot,\cdot;\ve \theta) \right).
    \label{GPprior}
\end{equation}
where $k(\cdot, \cdot; \ve \theta)$ is a covariance kernel and $\ve\theta$ is called hyperparameters. Furthermore, 
$\cov(f(\ve t),\ f(\ve t'))=k(\ve t, \ve t'; \ve \theta)$  for any  $ \ve t, \ve t' \in \R^d$. Let $\boldsymbol{f}_N=\left(f\left(\boldsymbol{t}_1\right), \ldots, f\left(\boldsymbol{t}_N\right)\right)^{\top}$. When the value of the hyper-parameters $\boldsymbol{\theta}$ is given, the posterior distribution, $p\left(\boldsymbol{f}_N \mid \mathcal{D}, \sigma^2\right)$, is a multivariate normal distribution with
$$
\begin{aligned}
\mathrm{E}\left(\boldsymbol{f}_N \mid \mathcal{D}, \sigma^2\right) & =\boldsymbol{C}_N\left(\boldsymbol{C}_N+\sigma^2 \boldsymbol{I}_N\right)^{-1} \ve Y_N, \\
\operatorname{Cov}\left(\boldsymbol{f}_N \mid \mathcal{D}, \sigma^2\right) & =\sigma^2 \boldsymbol{C}_N\left(\boldsymbol{C}_N+\sigma^2 \boldsymbol{I}_N\right)^{-1},
\end{aligned}
$$
where $\boldsymbol{Y}_N=(Y_1, \ldots,Y_N)^{\top}$, $\boldsymbol{I}_N$ is the $N\times N$ identity matrix,  covariance matrix $\boldsymbol{C}_N$ is calculated by using the kernel covariance function. Its $(i, l)$-th element is calculated by
$$
\boldsymbol{C}_N(i, l)=\operatorname{Cov}\left(f(\ve t_i), f(\ve t_l)\right)=k\left(\boldsymbol{t}_i, \boldsymbol{t}_l ; \boldsymbol{\theta}\right).
$$

To estimate $f_j(\cdot)$ in the nonlinear structured latent factor model (\ref{GPSIMFA}), we assume a  GP  priori \eqref{GPprior}, and use squared exponential kernel, that is, \begin{equation}
\boldsymbol{C}_{j}(i, l)=\operatorname{Cov}\left(f_j\left(\ve a_j^{\top} \ve x_i\right), f_j\left(\ve a_j^{\top} \ve x_l\right)\right)=\tau \exp \left(-\frac{w \left(\ve a_j^{\top} \ve x_i-\ve a_j^{\top} \ve x_l\right)^2}{2}\right).
\label{coveq1}
\end{equation}
Other types of covariance kernels can be found in   \cite{Rasmussen:2006} and \cite{Shi:2011}. We will assume the identifiability  assumptions are satisfied in the remaining of the paper. 


\subsubsection{Joint estimation of  loading matrix, factor scores and  hyperparameters}
Under the Bayesian analysis of NSLFA model, there are four types of parameters:
$\{\boldsymbol{a}_j,\ j=1, \ldots, J\}$ for loading coefficients, $\{\ve x_i,\ i=1, \ldots, N\}$ for factor scores, $\sigma^2$ for independent errors and $\{w,\tau\}$ involved in squared exponential covariance function \eqref{coveq1}. 
For convenience in notation, denote $\ve \theta \triangleq \{w,\tau,\sigma^2\}$ and $\ve \Theta$ represents all parameters. 
Recall that $\left(\boldsymbol{X}, \boldsymbol{A}\right)$ are the matrix forms of  factor scores
 $\ve x_{i}=(x_{i1},\ldots, x_{iK})^{\top}$ and loading coefficients $\boldsymbol{a}_{j}=(a_{j1},\ldots, a_{jK})^{\top}$. 

From the NSLFA model (\ref{GPSIMFA}) and the definition (\ref{GPprior}) of GP prior, the marginal distribution of $\ve Y_{j}=(Y_{1j}, \ldots, Y_{Nj})^{\top}$ given $\ve \Theta$ is obtained as a multivariate normal distribution,
\begin{equation}
    \ve Y_{j} \mid\ve \Theta \sim N\left(\ve{0}_N, \sigma^2 \boldsymbol{I}_N+\boldsymbol{C}_{j}\right).
    \label{distributionYn}
\end{equation}
Then the joint posterior distribution of $(\ve X, \ve A, \ve \theta)$ is calculated as follows\citep{Lawrence:2005}: 
\begin{eqnarray}
p(\ve X, \ve A, \ve \theta|\ve Y)  \propto \prod_{j=1}^{J}p\left(\boldsymbol{Y}_{j}| \boldsymbol{X}, \ve A, \ve \theta\right) p(\ve X) p(\ve \theta).
\notag
\end{eqnarray}
Therefore, the maximum a posterior distribution(MAP) estimates of $(\ve X, \ve A, \ve \theta)$ can be obtain through maximizing the posterior density function, or the following log density function:
\begin{equation}
    \ell_F(\mat{X},\ve{A},\ve \theta) = \sum_{j=1}^J {\left[  -\frac{1}{2} \log \abs{\sigma^2\boldsymbol{I}_N+\boldsymbol{C}_{j}}  -\frac{1}{2} \tr {\left((\sigma^2\boldsymbol{I}_N+\boldsymbol{C}_{j})^{-1} \vesub{Y}{j} \vesub{Y}{j}^\T  \right)} \right]} + \log p(\boldsymbol{X}).
    \label{jntlik}
\end{equation}
Furthermore, in order to address identifiability, we need to impose the constraints on $\ve{a}_j \in \mathcal{D}_j,\ j=1,\ \ldots, J$, where $\mathcal{D}_j=\left\{\ve{a}_j \in \mathbb{R}^K: a_{j k}=0\right.$ if $\left.q_{j k}=0\right\}$.
The optimization of maximization of \eqref{jntlik} requires the first derivatives of the objective function with respect to the unknown parameters, which are given in Appendix \ref{derivatives}.

Maximizing the log-likelihood function \eqref{jntlik} directly to calculate the joint estimate of these four types of unknown parameters may lead to some numerical problems.
We give an iterative estimation approach to solve this problem in the following
subsections

\subsubsection{An iterative estimation approach}

From model \eqref{GPSIMFA}, data are independent for $i=1, \ldots, N$ when given $\ve X, \ve A$ and the unknown function, i.e., $Y_{ij}|f_j(\cdot), \ve A, \ve X\sim N\left(f_j(\ve a_j^{\top} \ve x_i), \sigma^2\right)$ independently for $i=1, \ldots, N.$ The iterative approach starts with giving initial values of $\ve A$ and $\ve X$.

\textbf{Step 1}: Estimate $\ve \theta=(w, \tau, \sigma^2)$ given $\ve A$ and $\ve X$. This is similar to the Empirical Bayesian approach discussed in \cite{Shi:2011}. Specifically, 
we  maximize log likelihood function of the marginal distribution of $Y_{ij}$ to estimate $\ve \theta=(w, \tau, \sigma^2)$, 
    \begin{equation}
   \ell(\ve \theta|\ve Y, \ve A, \ve X) = \sum_{j=1}^J {\left[ - \frac{N}{2} \log(2\pi) -\frac{1}{2} \log \abs{\sigma^2\boldsymbol{I}_N+\boldsymbol{C}_{j}}  -\frac{1}{2} \tr {\left((\sigma^2\boldsymbol{I}_N+\boldsymbol{C}_{j})^{-1} \vesub{Y}{j} \vesub{Y}{j}^\T  \right)} \right]} + \log p(\boldsymbol{X}).
   \notag
\end{equation}
Once $\ve \theta$ is estimated, we can calculate the conditional distribution of $f_j(\cdot)$.
Let $t_j^*=\ve a_j^{\top} \ve x^*$ and  $t_{ij}=\ve a_j^{\top} \ve x_i$.  \cite{Shi:2011}(p.19, formula (2.7)) gives the explicit analytic expression of the prediction of $f_j\left(t_j^*\right)$ at the data point $t_j^*$ as follows \begin{equation}
    \mathrm{E}\left(f_j\left(t_j^*\right) \mid \mathcal{D}\right)=\boldsymbol{\psi}_j^{\top}(t_j^*)\left(\boldsymbol{C}_{j}+\sigma^2 \boldsymbol{I}_N\right)^{-1}\ve Y_j,
    \label{Efj}
\end{equation}
where $\boldsymbol{\psi}_j\left(t_j^*\right)=\left(k\left(t_j^*, t_{1j}\right), \cdots, k\left(t_j^*, t_{Nj}\right)\right)^{\top}$ is the covariance between $f_j\left(t_j^*\right)$ and $\boldsymbol{f}_{j}=\left(f_j\left(t_{1j}\right), \ldots, f_j\left(t_{Nj}\right)\right)^{\top}$.

\textbf{Step 2}: Estimate $\ve a_j$ and $\ve x_i$ given $f_j(\cdot)$, i.e., $\ve A$ and $\ve X$ separately for different $i$. Due to $Y_{ij}|f_j(\cdot)\sim N\left(f_j(\ve a_j^{\top} \ve x_i), \sigma^2\right)$ independently for $i=1, \ldots, N$, we can give the estimators for each $i$ separately.
For $i=1, \ldots, N,\ \ve x_i \in \underset{\ve x}{\arg \min }-\ell_i(\ve x)$, 
    where $$\ell_i(\ve x)=\sum_{j=1}^J Y_{i j} f_j(\ve a_j^{\top} \ve x)-\frac{1}{2}(f_j(\ve a_j^{\top} \ve x))^2.$$
For $j=1, \ldots, J,\ \ve a_j \in \underset{\ve a}{\arg \min }-\ell_j(\ve a)$, 
    where   $$\ell_j(\ve a)=\sum_{i=1}^N Y_{i j} f_j(\boldsymbol{a}^{\top} \ve x_i)-\frac{1}{2}(f_j(\boldsymbol{a}^{\top} \ve x_i))^2.$$

\textbf{Step 3}: Repeat Step 1 and Step 2 until convergence.
 Both
 $\ve x_{i}=(x_{i1},\ldots, x_{iK})^{\top}$ and $\boldsymbol{a}_{j}=(a_{j1},\ldots, a_{jK})^{\top}$ in Step 2 are $K$ dimensional vectors. Because $K$ is relatively small, this optimization algorithm is easy to implement. 
 
By the iterative algorithm, we can obtain the estimator of all parameters, then substituting those into Equation \eqref{Efj} to obtain an estimation of the unknown link function. More specifically,
  we use the  mean vector $\boldsymbol{\mu}_{j} =\boldsymbol{C}_{j}\left(\boldsymbol{C}_{j}+\sigma^2 \boldsymbol{I}_N\right)^{-1} \ve Y_{j}$ as the estimator of $\boldsymbol{f}_{j}$.

We  use the values of principle components in PCA as the initial values of $\ve X$ and uniform prior for $\ve A$ and $\ve \theta$. Ultimate value of $\ve X, \ve A, \ve \theta$ are estimated via the Scaled Conjungate Gradient(SCG) algorithm \citep{moller1993scaled}, we also use an easy optimization algorithm:  gradient descent algorithm as an alternative, and we can get the same result although gradient descent algorithm has low speed of convergence.

\section{Asymptotic Properties}
\label{identifiability}

In this section, we present the asymptotic properties for the estimators of parameters and unknown link function. Theorem 2 shows that the MAP estimators $\widehat{\ve a}_j^{\top}\widehat{ \ve x_i} 
 $ converge to the true value of parameter $\boldsymbol{a}_{j*}^{\top} \ve x_{i*}$ in probability by the general result of the consistency of the maximum-likelihood estimator in the discrete parameter stochastic processes in \cite{rao1980asymptotic} and each latent factor is
 consistently estimated if  satisfied the conditions of Theorem 1. Theorem 3 provides posterior consistency for $\left(f_j, \sigma^2\right)$ given the MAP estimators.

\textbf{Theorem 2.}  Let $\left(\ve X^*, \ve A^*\right) \in \mathcal{S}_Q$, where $\mathcal{S}_Q$ is defined in the Appendix \ref{A1A2}, be the true factor scores and loading matrix,
 $\ve x_{i*}$ is the $i$-th row of  $\ve X^*$ and $\ve a_{j*}$ is the $j$-th row of  $\ve A^*$. Let $\tau_*, \theta_*$ be the true values of hyperparameters.  Let $\left(\widehat{\ve a}_j^{T}\widehat{ \ve x_i}, \widehat{\tau}, \widehat{w}\right)$ denote the MAP estimators of $\left(\boldsymbol{a}_{j*}^{T}\ve x_{i*}, \tau_*, \theta_*\right)$. Under  Assumptions A1-A4, the MAP estimators are consistent estimators. That is, for $j=1,\ldots,J$,
\begin{equation}
\widehat{\ve a}_j^{\top}\widehat{ \ve x_i} \stackrel{p}{\rightarrow} \boldsymbol{a}_{j*}^{\top} \ve x_{i*},\quad \widehat{\tau} \stackrel{p}{\rightarrow} \tau_* \text { and } \widehat{w} \stackrel{p}{\rightarrow} w_* \ \text { as } N \rightarrow \infty.
\label{Thm2.1}
\end{equation}

Moreover, if $Q$ satisfies expression \eqref{Thm1} and thus the $k$-th latent factor  is structurally identifiable, then 
\begin{equation}
\E\sin \angle\left(\ve X_{[1: N, k]}^*, \widehat{\ve X}_{[1: N, k]}\right) {\rightarrow} 0\ \text{as}\ N{\rightarrow}\infty.
\label{Thm2.2}
\end{equation}

\textbf{Theorem 3.}  Let $P_0$ denote the joint conditional distribution of $\left\{Y_{ij}\right\}_{i=1}^{\infty}$ given the true factor scores $\ve X^*$ and loading matrix $\ve A^*$. Assuming that $f_{j*}$ is the true link,  $\sigma_*^2$ is the true variance of noise. Under Assumptions A1-A4, for every $\epsilon>0$,
$$
\Pi\left\{(f_j, \sigma) \in W_\epsilon \mid \boldsymbol{Y},  \widehat{\ve a}_j^{\top}\widehat{ \ve x_i}, \sigma_*^2, \widehat{\tau}, \widehat{w}\right\} \rightarrow 1 \text { in } P_0^N-\text { probability },
$$
where
$$
W_\epsilon=\left\{(f_j, \sigma): \left|f_j\left(\widehat{\ve a}_j^{\top}\widehat{ \ve x_i}\right)-f_{j*}\left(\boldsymbol{a}_{j*}^{\top} \ve x_{i*}\right)\right| <\epsilon,\left|\frac{\sigma}{\sigma_*}-1\right|<\epsilon\right\}.
$$
In other words, for each $j$,  we have 
$f_j\left(\widehat{\ve a}_j^{\top}\widehat{ \ve x_i}\right) \stackrel{p}{\rightarrow} f_{j*}\left(\boldsymbol{a}_{j*}^{\top} \ve x_{i*}\right) $  as $N  \to +\infty$.

The proof of both theorems are given in Appendix \ref{pf1}.

\section{Numerical results}
\label{Numerical}
Numerical illustrations of the NSLFA model are made based on the implementation procedure
 described before. For simulated examples in Section \ref{simulation}, we consider two scenarios, one is that all the latent factors are  identifiable and the other is that  a latent factor is not identifiable.
The ‘multi-phase oil flow’ data is discussed as an illustration of a real-world example in Section \ref{oil}. More simulated  examples are provided in Appendix \ref{Simulated}.

\subsection{Simulated Examples}
\label{simulation}
The true model to generate the data is Model \eqref{GPSIMFA} with the following setting 
$$
f_j(\boldsymbol{a}_j^{\top} \ve x)=\frac{1}{1+e^{-\boldsymbol{a}_j^{\top} \ve x}},   \quad y_j=f_j(\boldsymbol{a}_j^{\top} \ve x)+\epsilon_j,  \quad j=1,\ldots, J,
$$
and $\epsilon_j \sim N\left(0, \sigma^2\right)$ where $\sigma^2=0.25$.
To investigate the conditions of identifiability, we consider two scenarios, where the first one satisfies conditions of identifiability, but the second one violates them.  Let $|\cdot|$ denote the cardinality of a set.

\textit{(i) Scenario 1}. Set the design matrix as
\begin{equation}
Q_1^{\top} =
\left( \begin{array}{cccccc}
1 & 1 & \cdots & 0 & 0  & \cdots\\
0 & 0 & \cdots & 1 & 1  & \cdots\\
\end{array} \right).
\notag
\end{equation}
All latent factors are structurally identifiable with this simple structure, the “simple structure” design  is a safe design. Under the simple structure design, each manifest variable is associated with one and
only one factor, where $|R_{Q_1}(\{k\})|=J / 2, k=1,  2$. Given $Q_1$, $\boldsymbol{a}_j=\left(a_{j1}, 0\right)^{\top}$ for $j=1, \ldots, J/2$ and $\boldsymbol{a}_j=\left(0, a_{j2}\right)^{\top}$ for $j=J/2+1, \ldots, J$. Scenario 2 will be discussed in the next part. For both design structures,  different values of $J$ are considered and we let $N=5 J$. Specifically, we consider $J=6, 10, 20, 30,\ldots,.$ The factor scores $ \ve x_{1}, \ve x_{2} $
and the loading coefficients $\ve{a}_j s$ are generated    iid from distributions over the ball $\left\{\ve x \in \mathbb{R}^2:\|\ve x\| \leq 2.5\right\}$ . 

We use Models \eqref{GPSIMFA} and \eqref{GPprior} (NSLFA) and the implementation method  discussed in Section 2 to analyse the data. In this Simulation Study, we mainly investigate the convergence on estimating the latent variables and the unknown function. The former is measured by correlation and $\sin$ value between the latent variables and their estimation. To validate the accuracy of the estimation of $\ve X\ve A$, we used the quantity 
$d_{\ve X\ve A}=\|\ve X_*\ve A_*-\widehat{\ve X}\widehat{\ve A}\|_F^2/NJ$, 
where $\|\ve A\|_F=\sqrt{\sum_i \sum_j a_{i j}^2}$ denotes the Frobenius norm on the matrix space and $NJ$ is the total number of the elements of $\ve X_*\ve A_*$. To measure the convergence of the estimation of the unknown function $f_j$ for $j=1, \ldots, J$, we used the quantity $d_f=\frac{\sum_{j=1}^J\|\widehat{\ve f}_{j}-\ve f_{j*}\|^2}{NJ}$. For each sample size,  100  replications are conducted and  we take the average of those measures. For the sake of simplicity and convenience of notation, we denote ${\rm Corr}_{\ve x_1}=\corr({\ve x_1, \widehat{\ve x}_1)}$, ${\rm Corr}_{\ve x_2}=\corr({\ve x_2, \widehat{\ve x}_2)}$, ${\rm Sin}_{\ve x_1}=\sin({\ve x_1, \widehat{\ve x}_1)}$ and ${\rm Sin}_{\ve x_2}=\sin({\ve x_2, \widehat{\ve x}_2)}$.

\begin{table}[ht]
\renewcommand\arraystretch{0.9}
  \setlength{\tabcolsep}{1.5mm}
	\centering
	\begin{tabular}{@{}cccccccc@{}} 	\toprule 
	 &   & ${\rm Corr}_{\ve x_1}$ & ${\rm Corr}_{\ve x_2}$& ${\rm Sin}_{\ve x_1}$ & ${\rm Sin}_{\ve x_2}$ & $d_{\ve X\ve A}$ & $d_f$ \\  
	\midrule
 \multirow{4}{*}{\textit{NSLFA}}
  & \textit{J=6} &  0.88& 0.90 &      0.38& 0.33 &  2.47& 0.578 \\
  & \textit{J=10} & 0.92 &0.92 &      0.35 &0.32& 2.02 &0.453\\
  &\textit{J=20}  & 0.98 &0.98 &     0.21& 0.22& 0.98 &0.332\\
   &\textit{J=30}   & 0.99& 0.99&   0.15&0.14    &  0.54 & 0.231\\

   \midrule
\multirow{6}{*}{\textit{Linear FA}} 
& \textit{J=6} &  0.78& 0.74 &      0.61& 0.65 &  10.25&  \\
  & \textit{J=10} & 0.84 &0.80 &     0.52 &0.58& 7.47&\\
  &\textit{J=20}  & 0.89 &0.86 &     0.43& 0.49& 4.21 &\\
   &\textit{J=30}   & 0.91& 0.88 &   0.41 &0.46    & 2.70 & \\
    &\textit{J=40}   & 0.91& 0.89 &   0.39 &0.44    & 2.70 & \\
   & \textit{J=50}  &0.92&0.90 &  0.37 &0.42 & 1.08 &\\
 \midrule
   \multirow{6}{*}{\textit{GPLVM}} 
& \textit{J=6} &  0.32& 0.18 &      0.94& 0.98&  & 0.325  \\
  & \textit{J=10} & 0.72 &0.67 &     0.69 &0.73& &0.254\\
  &\textit{J=20}  & 0.79 &0.72 &     0.60& 0.69&  &0.161\\
   &\textit{J=30}   & 0.80& 0.50 &   0.75 &0.72    & &0.031 \\
   & \textit{J=40}  &0.85&0.44 &  0.52 &0.89 & &0.025\\
    & \textit{J=50}  &0.85&0.27 &  0.52 &0.96 & &0.010\\

 \bottomrule 
	\end{tabular}
\vspace{-1mm}
\caption{The result of \textit{NSLFA}, \textit{Linear FA} and \textit{GPLVM} in Scenario 1}
\label{Example1}
\end{table}

\begin{table}[ht]
\renewcommand\arraystretch{0.9}
  \setlength{\tabcolsep}{1.50mm}
  \caption{The result of  \textit{Scenario 2} using NSLFA}
	\centering
	\begin{tabular}{@{}cccccccc@{}} 	\toprule 
	&   & ${\rm Corr}_{\ve x_1}$ & ${\rm Corr}_{\ve x_2}$& ${\rm Sin}_{\ve x_1}$ & ${\rm Sin}_{\ve x_2}$ & $d_{\ve X\ve A}$ & $d_f$ \\   
	\midrule

\multirow{5}{*} 
& \textit{J=5} &   0.88& 0.43 &     0.38& 0.88 &  3.75&0.820\\
 & \textit{J=10} & 0.92 &0.43 &      0.35 &0.86& 3.12&0.675 \\
  &\textit{J=20}  & 0.98& 0.47 &     0.21& 0.84& 2.26 &0.451\\
   &\textit{J=30}   &0.99 &0.49 &  0.15& 0.81    & 1.37 &0.235\\
   & \textit{J=50}  & 0.99 &0.56 &     0.03& 0.78  & 0.09&0.110\\

 \bottomrule 
	\end{tabular}
\vspace{-1mm}
\label{Example1S2}
\end{table}

We use two gradient-based algorithm respectively in the step of optimization,   both scale conjugate gradient algorithm and gradient descent algorithm lead to the same result although they have different convergence speed. 

\begin{figure}[h]
  \centering
  \includegraphics[width=15cm,height=8cm]{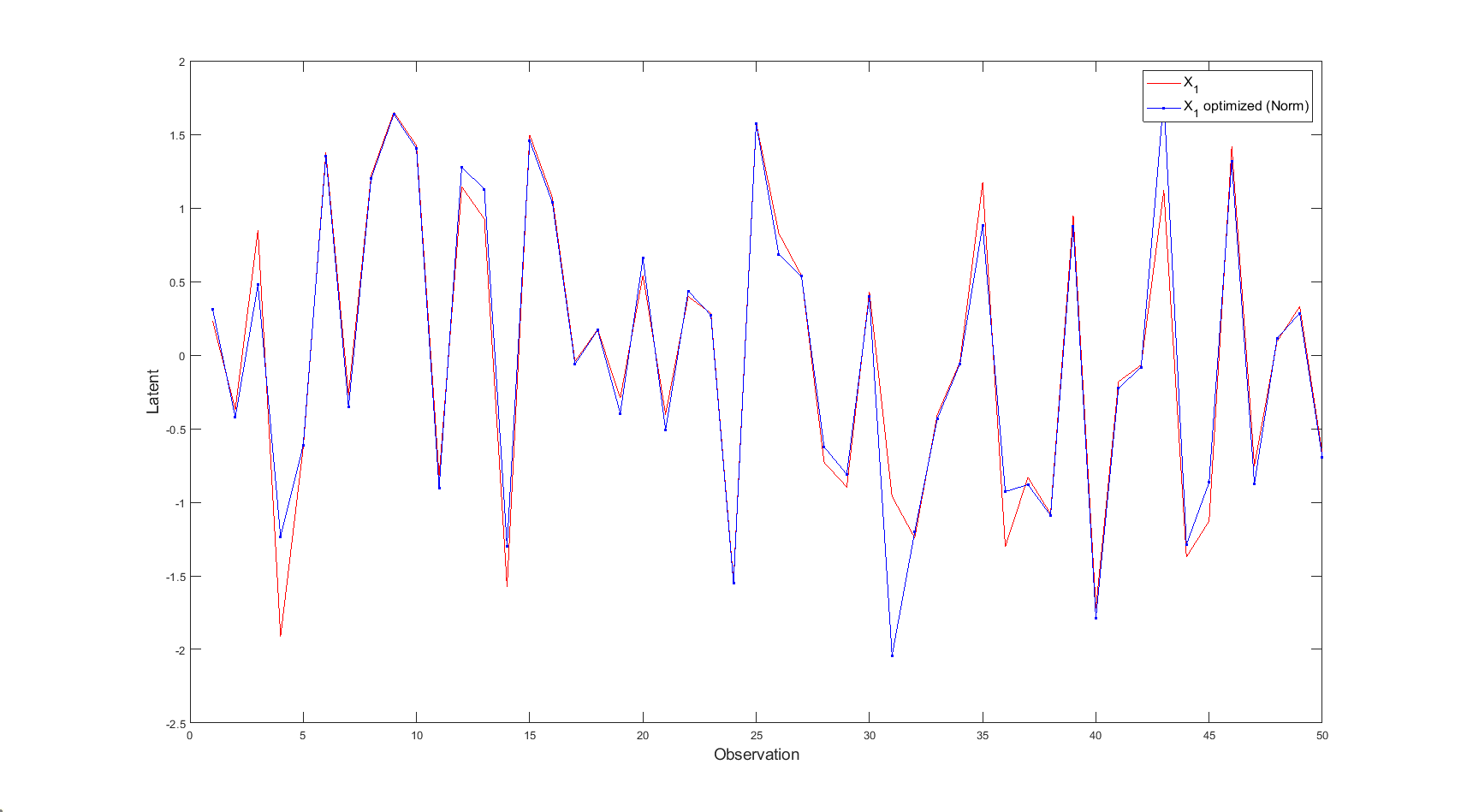}
  \caption{The true value and estimation of $x_1$ by using NSLFA in Scenario 1.}
  \label{fig:x1}
\end{figure}

\begin{figure}[h]
  \centering
  \includegraphics[width=15cm,height=8cm]{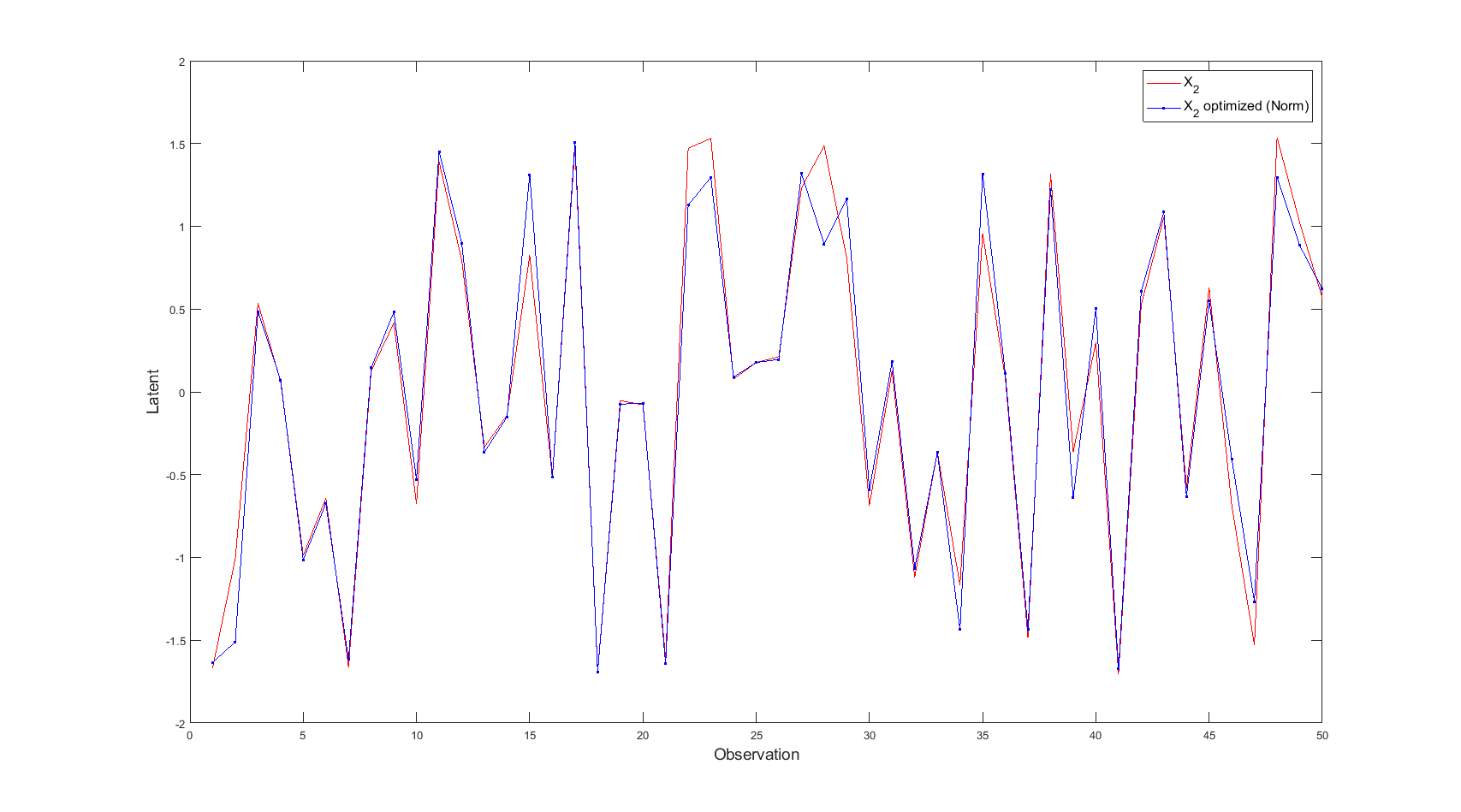}
  \caption{The true value and estimation of $x_2$ by using NSLFA in Scenario 1.}
  \label{fig:x2}
\end{figure}

As comparison, we also use a linear factor analysis model (denoted by LFA) and GPLVM \citep{Lawrence:2005}. In LFA, we use the same constraints applied to the loading matrix while there are no  constraints that can be imposed on GPLVM. The results of the simulation study are presented in Table \ref{Example1} and Figures \ref{fig:x1}-\ref{fig:x2}. 

\begin{figure}[h]
  \centering
  \includegraphics[width=15cm,height=8cm]{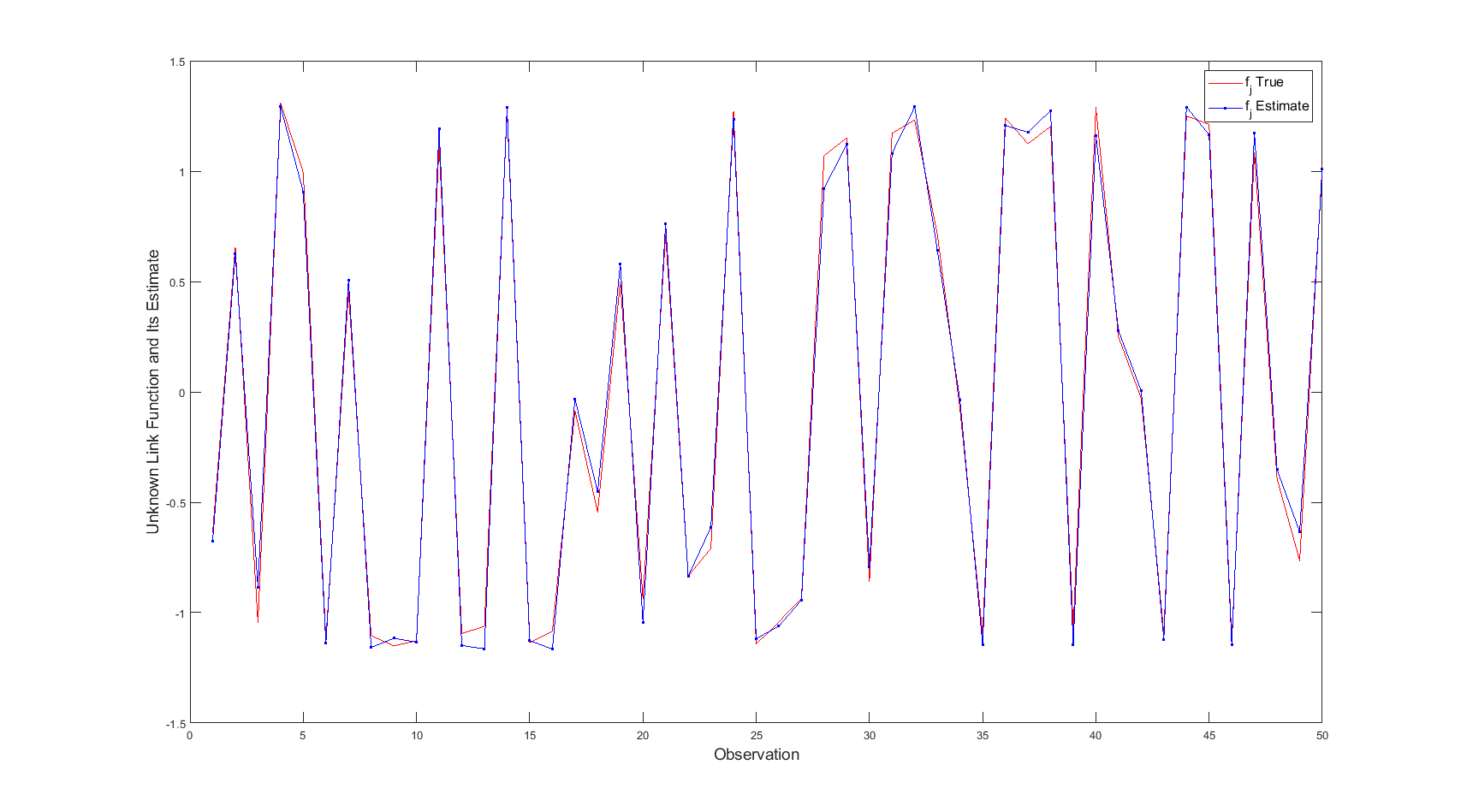}
  \caption{The true value and estimation of $f_j$ by using NSLFA in Scenario 1.}
  \label{fig:fj}
\end{figure}

Figures \ref{fig:x1} and \ref{fig:x2} are the true value and estimation of $x_1$ and $x_2$  evaluated from 50 observations when $J=10$ and $N=50$ in Scenario 1 for one replication using NSLFA. For this replication, the estimation of $x_1$ is  a good estimate of the true $x_1$. In this case, the correlation between the true value and the estimate of $x_1$ is  0.92 and the value of sin is 0.35 based on 100 replications. The correlation is close to 1, indicating
the identifiability of the estimated factor scores. It is also confirmed by the small $\sin$ values. The performance of the estimation for $\ve x_2$ is very similar in this case.
The convergence of the estimation of the factor scores is mainly dependent on $J$. 
Table \ref{Example1} shows the results with different values of $J$.
It shows clearly  that the correlation tends to 1 and the $\sin$ value tends to 0 as $J$ is increasing. The accuracy on estimating unknown parameters involved in $A$ and the unknown function $f$ depends on both $N$ and $J$. 
The former is measured by $d_{\ve X\ve A}$ while the latter is measured by $d_f$. The results reported in Table \ref{Example1} show the good accuracy of the estimation for both groups of unknown quantities, and improved as $N$ and $J$ are increasing. Figure \ref{fig:fj} shows the true value of $f$ and its estimated value for one replication with $J=10$ and $N=50$. 

Comparing the results by using LFA, we can see the accuracy on estimating both the factor scores and the parameters is not comparable with the proposed model NSLFA due to its ignorance nonlinearity as expected.

Both NSLFA and GPLVM have a good estimation  of the
unknown function. Even if
the latent space lacks uniqueness, GPLVM still can  extract meaningful information and capturing patterns in the data, leading
to effective prediction performance. However, there is a lack of stability in accurately estimating the factor scores of the GPLVM model. The lack of stability is mainly attributed to the oversight of of the identifiability of the factor scores.  Ignoring identifiability can lead to multiple plausible solutions for the factor scores, resulting in unstable estimates.

\textit{(ii) Scenario 2: a simulated example without satisfing identifiability conditions}.
We use the same settings as those discussed in the previous part except the structure of the design matrix, where  in  Scenario 2,
\begin{equation}
Q_2^{\top} =
\left( \begin{array}{cccccc}
1 & 1 & \cdots & 1 & 1  & \cdots\\
0 & 0 & \cdots & 1 & 1  & \cdots\\
\end{array} \right).
\notag
\end{equation}
The purpose of  Scenario 2 in the simulation example is to investigate the conditions of identifiability. From the setting and Theorem 1, the factor $x_1$ is identifiable but $x_2$ is not. 
Given $Q_2$, $\boldsymbol{a}_j=\left(a_{j1}, 0\right)^{\top}$ for $j=1, \ldots, J/2$ and $\boldsymbol{a}_j=\left(a_{j1}, a_{j2}\right)^{\top}$ for $j=J/2+1, \ldots, J$,  $|R_{Q_2}(\{1\})|=1 / 2>0$, $|R_{Q_2}(\{1,2\})|=1 / 2>0$ and $R_{Q_2}(\{2\})$ is empty, when $k=2$,  
$$\bigcap_{2 \in S,\ R_{Q_2}(S)\ non-empty} S=\{1,2\}\neq\{2\},$$ so the second factor  is not identifiable. 

As we can see from Table \ref{Example1S2} that  convergence of $\ve x_1$ is almost the same as in Scenario 1 and ${\rm Corr}_{\ve x_2}$ is not close to 1,  ${\rm Sin}_{\ve x_2}$ not close to 0 even for large $J$, so estimation of $\ve x_2$ is not identifiable. In linear factor analysis, rotation ambiguity leads to nonidentifiability in both loading coefficients and factor scores, while the product of the loadings and factor scores remains constant.  This is true for the  NSLFA model as well, indicating that the convergence of
$\ve X\ve A$ is not impacted. This is the fact shown in Table \ref{Example1S2}, we can  find the accuracy of the estimation of $\ve X\ve A$ is comparable with the results given in Table \ref{Example1} where the identifiability conditions are satisfied. The accuracy of the estimation improves as $NJ$ increases. The same phenomenon is found for the estimation of the unknown function. The accuracy of the estimation is not affected by the  identifiability conditions and is improving when the sample size is increasing.
This is because 
even if the latent space lacks uniqueness, factor analysis and other related models like GPLVM excel at extracting meaningful information and capturing patterns in the data, leading to effective prediction performance.

We have conducted simulation studies with different settings. Appendix \ref{Simulated} presented the results with $K=3$ and $K=5$ factors. We have the findings similar to the above.


\clearpage

\subsection{Real Data Analysis}
\label{oil}

We use the ‘multi-phase oil flow’
data \citep{bishop1993analysis} to illustrate the usefulness of NSLFA model for a real world example. This is synthetic data modelling non-intrusive measurements on a pipe-line transporting a mixture of oil, water and gas. The flow in the pipe takes one out of three possible configurations: horizontally stratified, nested annular or homogeneous mixture flow. The data are collected in a 12-dimensional measurement space, but for each configuration, there is only two degrees of freedom: the fraction of water and the fraction of oil. The fraction of gas is redundant, since the three fractions must sum to one. Hence, the data can be locally  approximated 2-dimensional.
The data set is artificially generated and therefore is known to lie on a lower dimensional manifold. For more details on the data see: https://inverseprobability.com/3PhaseData.html.
Here we use a sub-sampled version of the data, containing 100 data points, to demonstrate the fitting
of NSLFA model and the usefulness of the nonlinear model involved.  

We conducted linear factor analysis on the data and used the varimax method to automatically set some coefficients to 0, in order to meet the identifiability conditions. After that, we compared our model with the linear factor analysis.
In Figure \ref{oil100} we show  the visualisation obtained using the LFA and NSLFA model. The latent-space of NSLFA demonstrates significantly better results in terms of separating different flow phases compared to those obtained by the LFA model. To assess the quality of the visualizations objectively, we classified each data point based on the class of its nearest neighbor in the two-dimensional latent-space provided by each method. The classification error for LFA is 28, while for NSLFA it is only 7.

\clearpage
\begin{landscape}
\newgeometry{left=3.5cm,right=2cm,top=10cm}
\begin{figure}[htbp]
\centering
\includegraphics[height=10cm,width=25cm]{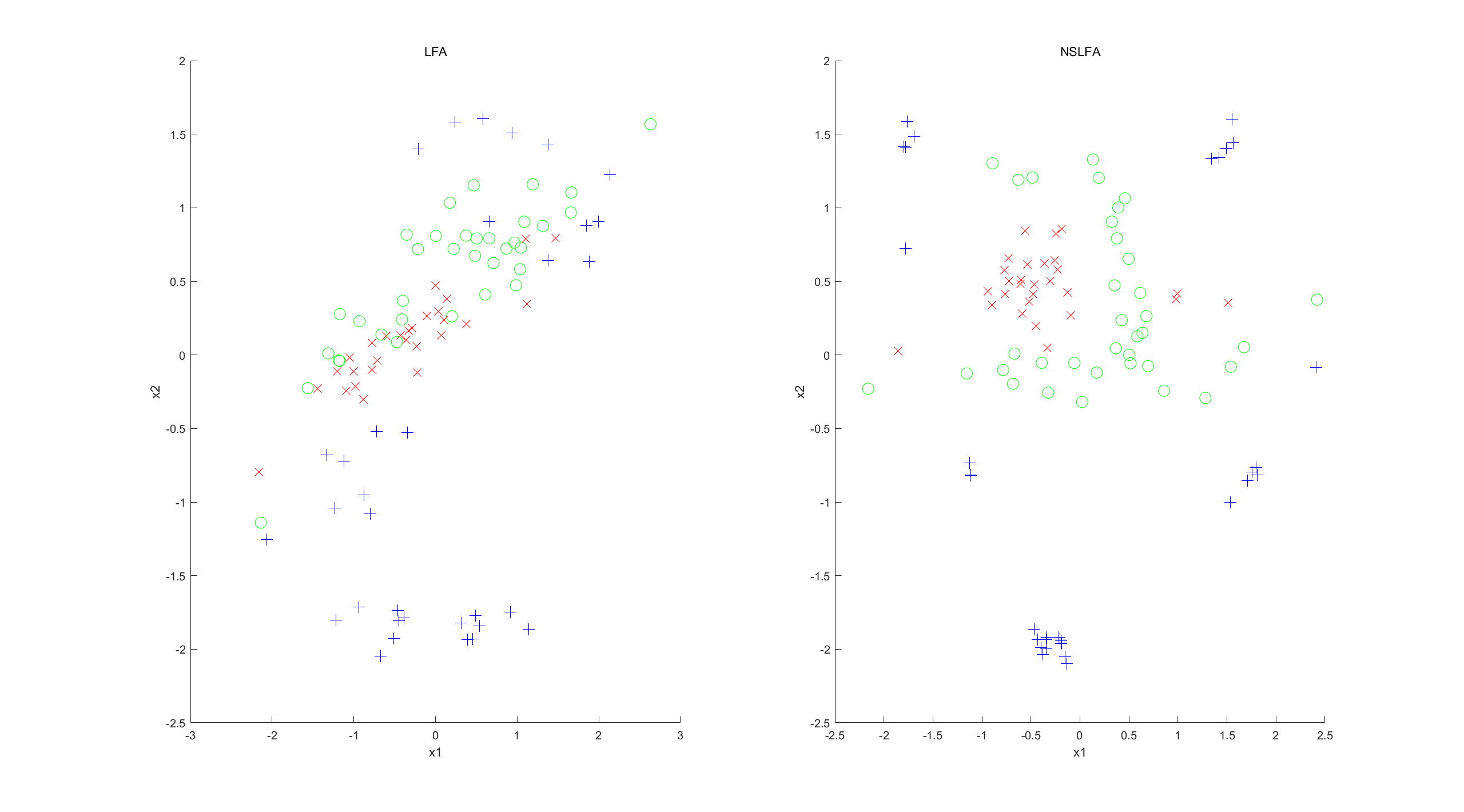}\\
\caption{Visualisation of the Oil data with LFA(left) and  NSLFA(right).}
\label{oil100}
\end{figure}
\end{landscape}
\restoregeometry

\section{Discussion}
\label{concluding}
In this paper,  we  develope a nonlinear structured latent factor analysis model that is capable of identifying and interpreting factors even when dealing with nonlinear data. We study how design information affects the identifiability and estimability of NSLAF model. We give necessary and sufficient conditions to achieve the structural identifiability of a given latent
factor.  We give an  iterative two-step approach to recover the structurally identifiable latent factors
 and  efficiently estimate the unknown function. In simulation study, our NSLAF model performs better than both LFA model and GPLVM. And we also demonstrate the model on a real-world  data set.

 The proposed NSLFA models \eqref{GPSIMFA} and \eqref{GPprior} are an extension of high dimensional single index models \citep{radchenko2015high} to a latent variable model. A more general model is to replace \eqref{GPSIMFA} by $Y_j=f_j(x_1, \ldots, x_K)+\varepsilon_j$, where $f_j$ is unknown. This method is similar to the GPLVM \citep{Lawrence:2005, damianou2021multi,lalchand2022generalised}, which is popular in machine learning and other areas. But the identifiability problem is usually ignored, which is not a problem when the model is used in prediction or classification, but the nonidentifiability limits the interpretation  of the latent scores and then limits the application. Research along this direction is carrying on.

\newpage


\newpage
\bibliographystyle{apalike}
\bibliography{thesisbiblio}

\newpage
\appendix

 \section{Multivariate linear regression model }
To help understanding, we put the  definition of the multivariate linear regression model is
$$
\boldsymbol{y}_i=\boldsymbol{A}^{\top} \boldsymbol{x}_i+\boldsymbol{\epsilon}_i
$$
for $i=1, \ldots, N$, where the response variables $\ve y_i=(Y_1, \ldots, Y_J)^\top$ and $J\geq2$, the predictor variables $\ve x_i=(x_1, x_2, \ldots, x_K)^\top$. So we can define the GP linear FA model  as (too difficult, need to consider the correlation in $\boldsymbol{f}$): 
$$\boldsymbol{y}_i=\boldsymbol{f}(\boldsymbol{A}^{\top} \boldsymbol{x}_i+\boldsymbol{\epsilon}_i)$$

For details, we can write $$\boldsymbol{Z} = \boldsymbol{X}\boldsymbol{A} + \boldsymbol{E},$$ 

where  the $N \times J$ matrix

$$\boldsymbol{Z}=\left[\begin{array}{ccccc}
Y_{1,1} & Y_{1,2} & \ldots & Y_{1, J} \\
Y_{2,1} & Y_{2,2} & \ldots & Y_{2, J} \\
\vdots & \vdots & \ddots & \vdots \\
Y_{N, 1} & Y_{N, 2} & \ldots & Y_{N, J}
\end{array}\right]=\left[\begin{array}{llll}
\boldsymbol{Y}_1 & \boldsymbol{Y}_2 & \ldots & \boldsymbol{Y}_J
\end{array}\right]=\left[\begin{array}{c}
\boldsymbol{y}_1^{\top} \\
\vdots \\
\boldsymbol{y}_N^{\top}
\end{array}\right].$$

The $N \times K$ design matrix of predictor variables is
$$
\boldsymbol{X}=\left[\begin{array}{cccc}
x_{1,1} & x_{1,2} & \ldots & x_{1, K} \\
x_{2,1} & x_{2,2} & \ldots & x_{2, K} \\
\vdots & \vdots & \ddots & \vdots \\
x_{N, 1} & x_{N, 2} & \ldots & x_{N, K}
\end{array}\right]=\left[\begin{array}{llll}
\boldsymbol{v}_1 & \boldsymbol{v}_2 & \ldots & \boldsymbol{v}_K
\end{array}\right]=\left[\begin{array}{c}
\boldsymbol{x}_1^{\top} \\
\vdots \\
\boldsymbol{x}_N^{\top}
\end{array}\right].
$$

The $K \times J$ matrix
$$\boldsymbol{A}=\left[\begin{array}{cccc}
a_{1,1} & a_{1,2} & \ldots & a_{1, J} \\
a_{2,1} & a_{2,2} & \ldots & a_{2, J} \\
\vdots & \vdots & \ddots & \vdots \\
a_{K, 1} & a_{K, 2} & \ldots & a_{K, J}
\end{array}\right]=\left[\begin{array}{llll}
\boldsymbol{a}_1 & \boldsymbol{a}_2 & \ldots & \boldsymbol{a}_J
\end{array}\right].$$

The $N \times J$ matrix
$$\boldsymbol{E}=\left[\begin{array}{cccc}
\epsilon_{1,1} & \epsilon_{1,2} & \ldots & \epsilon_{1, J} \\
\epsilon_{2,1} & \epsilon_{2,2} & \ldots & \epsilon_{2, J} \\
\vdots & \vdots & \ddots & \vdots \\
\epsilon_{N, 1} & \epsilon_{N, 2} & \ldots & \epsilon_{N, J}
\end{array}\right]=\left[\begin{array}{ll}
\boldsymbol{e}_1 \ \boldsymbol{e}_2 \ldots \boldsymbol{e}_J
\end{array}\right]=\left[\begin{array}{c}
\boldsymbol{\epsilon}_1^{\top} \\
\vdots \\
\boldsymbol{\epsilon}_N^{\top}
\end{array}\right] .$$

\section{Log-likelihood derivatives}\label{derivatives}
\subsection{Hyperparameter first derivatives}

Let $\ve{\theta} = (w,\tau, \sigma^2 )$  be the vector of kernel hyperparameters. Let $\mat{K}_j=\sigma^2\boldsymbol{I}_N+\boldsymbol{C}_{j}$, the log-likelihood gradient is then given by
\begin{align}
\frac{\partial \ell(\mat{X},\ve{A}) }{\partial \theta}  
				&=  \sum_{j=1}^J \left[ -\frac{1}{2} \tr \left( \mat{K}_j^{-1} \frac{\partial \mat{K}_j}{\partial \theta} \right) + \frac{1}{2} \left( \mat{Y}_j^\T  \mat{K}_j^{-1} \frac{\partial \mat{K}_j}{\partial \theta} \mat{K}_j^{-1}  \mat{Y}_j\right) \right] \notag \\
				&=\sum_{j=1}^J \frac{1}{2} \tr \left( (\ve{\alpha_j} \ve{\alpha_j}^\T -\mat{K}_j^{-1} ) \frac{\partial \mat{K}_j}{\partial \theta}  \right), \quad \text{ where } \quad \ve{\alpha_j} = \mat{K}_j^{-1}  \mat{Y}_j.
    \notag
\end{align}
The derivatives $ \partial \mat{K}_j/\partial \theta  $ are $N \times N$ dimensional matrices as follows 
\begin{align}
\left( \frac{\partial \mat{K}_j}{\partial w}  \right)
&=  -\frac{1}{2} \mat{D}_{j1} \odot  (\mat{K}_j - \sigma^2 \mat{I}_N), \notag\\
\left( \frac{\partial \mat{K}_j}{\partial \tau} \right) &= \frac{1}{\tau} (\mat{K}_j - \sigma^2 \mat{I}_N),
\notag \\
\left( \frac{\partial \mat{K}_j}{\partial \sigma^2} \right)
&= \mat{I}_N,
\notag 
\end{align}
where $\odot$ represents the Hadamard (element-wise) product and $\mat{D}_{j1}$ is also an $N \times N$ matrix whose $il^{th}$ element is given by $(\ve a_j^{\top} \ve x_i-\ve a_j^{\top} \ve x_l)^2$.

 \subsection{Latent variables first derivatives}

The first derivatives of the log-likelihood with respect the new latent variables is as follows
\begin{align} 
c_{ik}= \frac{\partial \ell(\mat{X}, \ve{A}) }{\partial x_{ik}}
& =  \sum_{j=1}^{J} \left[-\frac{1}{2} \tr \left( \mat{K}_j^{-1} \frac{\partial \mat{K}_j}{\partial x_{ik}} \right) + \frac{1}{2} \left( \mat{Y}_j^\T  \mat{K}_j^{-1} \frac{\partial \mat{K}_j}{\partial x_{ik}} \mat{K}_j^{-1}  \mat{Y}_j \right) \right] \notag \\ 
& =  \sum_{j=1}^{J} \left[\frac{1}{2} \tr \left( (\ve{\alpha}_j \ve{\alpha}_j^\T -\mat{K}_j^{-1} ) \frac{\partial \mat{K}_j}{\partial x_{ik}}  \right) \right],\quad  \text{ where } \quad \ve{\alpha}_j = \mat{K}_j^{-1}  \mat{Y}_j .\notag
\end{align}

The $ \partial \mat{K}_{j}/\partial x_{ik} $ is an $N \times N$ sparse matrix of all zeros but the $i^{th}$ row and column; the $(i,i)^{th}$ position is also zero, that is
\begin{equation}
\left( \frac{\partial \mat{K}_{j}}{\partial x_{ik}} \right) =
\begin{pmatrix} 
  &  \cdots &  *      &  &\cdots  \\
  &  \vdots &  c_{li} &  &\vdots \\
  &  \vdots &  *	  &  &\vdots \\
* & c_{il}  &  0      &* & * & \\
  &  \vdots &  *	  &  &\vdots \\
  &  \cdots &  *      &  &\cdots \\\notag
\end{pmatrix}.
\end{equation}

The elements $c_{il}$ of the $i^{th}$ row/column are given by
\begin{equation*}
c_{il} = \left( \frac{\partial \mat{K}_{j}(i,l)}{\partial x_{ik}} \right) = 
 -w (\ve a_j^{\top} \ve x_i-\ve a_j^{\top} \ve x_l)a_{jk}\boldsymbol{C}_{j}(i, l)  .
\end{equation*}

Given the sparse structure of $\partial \mat{K}_j/\partial x_{ik} $ and the symmetry of $\ve{\alpha}_j \ve{\alpha}_j^\T -\mat{K}_j^{-1}$, then the trace calculation  simplifies somewhat. Let us see that with a specific example how to calculate $\tr (\mat{A} \mat{B}) $. The matrices $\mat{A}$ and $\mat{B}$ are symmetric and the latter is also sparse as already discussed. Then,  
\begin{align*}
& \tr \left [
\begin{pmatrix}
a_{11} & a_{12} & a_{13} & \cdots & a_{1N} \\
a_{21} & a_{22} & a_{23} & \cdots & a_{2N} \\
a_{31} & a_{32} & a_{33} & \cdots & a_{3N} \\
\vdots & 				&\vdots  &        & \vdots \\
a_{N1} & a_{N2} & a_{N3} & \cdots & a_{NN} \\
\end{pmatrix}
\begin{pmatrix}
0 & 0 & b_1 & \cdots &0 \\
0 & 0 & b_2 & \cdots &0 \\
b_1 & b_2 & 0 & \cdots &b_N \\
\vdots & & \vdots & & \vdots \\
0 & 0 & b_N & \cdots &0 \\
\end{pmatrix} \right ] \\
& = 
a_{13}b_1 + a_{23}b_2+ \left( a_{31}b_1 + a_{32}b_2 + 0 + \ldots +  a_{3N}b_N\right) + \ldots + a_{N3}b_N \\
& = 
(a_{13}+a_{31}) b_1 + (a_{23}+a_{32}) b_2 + 0 + \ldots +  (a_{3N}+ a_{N3}) b_N \\
& = 
2(a_{13} b_1 + a_{23} b_2 + 0 + \ldots +  a_{3N} b_N) .
\end{align*}

\subsection{Factor loadings first derivatives}
For those $a_{mk} \neq 0$ as indicated by design matrix $Q$, the first derivatives of the log-likelihood with respect the factor loadings is as follows
\begin{align} 
\frac{\partial \ell(\mat{X}, \ve{A}) }{\partial a_{mk}}
& =  \sum_{j=1}^{J} \left[-\frac{1}{2} \tr \left( \mat{K}_j^{-1} \frac{\partial \mat{K}_j}{\partial a_{mk}} \right) + \frac{1}{2} \left( \mat{Y}_j^\T  \mat{K}_j^{-1} \frac{\partial \mat{K}_j}{\partial a_{mk}} \mat{K}_j^{-1}  \mat{Y}_j \right) \right] \notag \\ 
& =  \sum_{j=1}^{J} \left[\frac{1}{2} \tr \left( (\ve{\alpha}_j \ve{\alpha}_j^\T -\mat{K}_j^{-1} ) \frac{\partial \mat{K}_j}{\partial a_{mk}}  \right) \right],\quad  \text{ where } \quad \ve{\alpha}_j = \mat{K}_j^{-1}  \mat{Y}_j .\notag
\end{align}

$\partial \mat{K}_{j}/\partial a_{mk} =\ve{0}$ if $j\neq m$ and $ \partial \mat{K}_{j}/\partial a_{jk} =  -\mat{D}_{j2} \odot  (\mat{K}_j - \sigma^2 \mat{I}_n),$ 
where $\odot$ represents the Hadamard (element-wise) product and $\mat{D}_{j2}$ is also an $n \times n$ matrix whose $il^{th}$ element is given by $w(\ve a_j^{\top} \ve x_i-\ve a_j^{\top} \ve x_l)(x_{ik}-x_{lk})$.

\section{Simulated examples}
\label{Simulated}
In this section, we present additional simulation studies, aiming to demonstrate the robustness of the conclusions derived from the simulation study. Here we consider $K=3$
and 5 and more manifest variables. We examine two scenarios: the first scenario satisfies the identifiability conditions, while the second scenario violates them, with all other conditions remaining consistent with the simulation study.

\textit{Setting 1}.
A simple structure is not
necessary for a good measurement design. A latent factor can
still be identified even when it is always measured together
with some other factors, we call this a mixed structure. In Setting 1, we consider $K=3$. We consider two scenarios,
where the first one satisfies conditions of identifiability, but the second one
violates them. To be specific, in the first scenario,
\begin{equation}
Q_1^{\top} =
\left( \begin{array}{cccccccc}
1 & 1 &  0& \cdots & 1 & 1 & 0 &\cdots\\
1 & 0 &  1 &\cdots & 1 & 0 & 1 &\cdots\\
0 & 1 &  1 &\cdots & 0 & 1 & 1 &\cdots\\
\end{array} \right),
\notag
\end{equation}
all latent factors are structurally identifiable even
when there is no item measuring a single latent factor, 
where $|R_{Q_1}(S)| = J/3, S = \{1, 2\},\{1, 3 \}\text{and} \{2,3\}.$ 
In the second scenario,
\begin{equation}
Q_2^{\top} =
\left( \begin{array}{cccccccc}
1 & 1 &  0& \cdots & 1 & 1 & 0 &\cdots\\
\textbf{0} & 0 &  1 &\cdots & \textbf{0} & 0 & 1 &\cdots\\
0 & 1 &  1 &\cdots & 0 & 1 & 1 &\cdots\\
\end{array} \right),
\notag
\end{equation}
the first and the third latent factors are identifiable, while the second latent factor is not identifiable, the design structure is given by  $|R_{Q_2}(S)| = J/3, S = \{1\},\{1, 3 \}\text{and} \{2,3\}.$ 
For both design structures above, a range of $J$ values are considered and we let $N=5J$. Specifically, we consider $J=6, 12, 21, 30,99, \ldots, 252$.

\begin{table}[ht]
\renewcommand\arraystretch{0.9}
  \setlength{\tabcolsep}{0.8mm}
	\centering
	\caption{The result of \textit{NSLFA},  \textit{Linear FA} and \textit{GPLVM} in Setting 1}
	\label{Example2}
	\begin{tabular}{@{}cccccccccc@{}} 	\toprule 
	 &   & ${\rm Corr}_{\ve x_1}$ & ${\rm Corr}_{\ve x_2}$& ${\rm Corr}_{\ve x_3}$& ${\rm Sin}_{\ve x_1}$ & ${\rm Sin}_{\ve x_2}$& ${\rm Sin}_{\ve x_3}$ &  $d_{\ve X\ve A}$ & $d_f$ \\  
	\midrule
 \multirow{7}{*}{\textit{NSLFA}}
  & \textit{J=6} &  0.82& 0.83 &      0.82&0.45& 0.44&0.45 &  7.77& 0.615\\
  & \textit{J=12} & 0.85 &0.88 &      0.86 &0.40& 0.39&0.41 &6.71&0.521\\
  &\textit{J=21}  & 0.90 &0.89 &      0.90&0.38& 0.38&0.38&6.45&0.410\\
   &\textit{J=30}   & 0.92 &0.91 &      0.92&0.34& 0.34&0.32 &3.56&0.398\\
   & \textit{J=99}  & 0.95& 0.95 &    0.93 &0.15  & 0.17&0.16&1.45&0.215\\
   & \textit{J=252}  & 0.99& 0.99 &    0.99 &0.15  & 0.12&0.10&0.50&0.175\\

   \midrule
   \multirow{7}{*}{\textit{Linear FA}}
      & \textit{J=6} &  0.60& 0.65 &      0.63&0.75& 0.80&0.77 &  25.45& \\
  & \textit{J=12} & 0.68 &0.65 &     0.68 &0.65& 0.68&0.67 &18.63&\\
  &\textit{J=21}  & 0.75 &0.73&      0.73&0.51& 0.56&0.49 &14.25&\\
   &\textit{J=30}   & 0.80 &0.78 &      0.79 &0.49& 0.45&0.42 &10.25&\\
   & \textit{J=99}  & 0.85& 0.85 &  0.84 &0.37 & 0.35&0.36&7.55&\\
   & \textit{J=252}  & 0.85& 0.87 &    0.86 &0.35  & 0.37&0.40&5.53&\\

   \midrule
   \multirow{5}{*}{\textit{GPLVM}}
      & \textit{J=6} &  0.50& 0.35 &      0.23&0.85& 0.98&0.99 &  & 0.580\\
  & \textit{J=12} & 0.58 &0.31 &     0.24&0.86& 0.99&0.99 &&0.451\\
  &\textit{J=21}  & 0.58 &0.31&      0.33&0.81& 0.96&0.99 &&0.336\\
   &\textit{J=30}   & 0.55 &0.14 &      0.33 &0.79& 0.99&0.95 &&0.221\\
   & \textit{J=60}  & 0.58& 0.45 &  0.22 &0.77 & 0.98&0.96&&0.120\\

 \bottomrule 
	\end{tabular}
\vspace{-1mm}
\end{table}

\begin{table}[ht]
\renewcommand\arraystretch{0.9}
  \setlength{\tabcolsep}{0.8mm}
	\centering
	\caption{The result of \textit{scenario 2} in Setting 1}
	\label{Example2S2}
	\begin{tabular}{@{}cccccccccc@{}} 	\toprule 
	 &   & ${\rm Corr}_{\ve x_1}$ & ${\rm Corr}_{\ve x_2}$& ${\rm Corr}_{\ve x_3}$& ${\rm Sin}_{\ve x_1}$ & ${\rm Sin}_{\ve x_2}$& ${\rm Sin}_{\ve x_3}$  &  $d_{\ve X\ve A}$ & $d_f$ \\
 \midrule
\multirow{7}{*}
  & \textit{J=6} &  0.82& 0.45 &      0.82&0.45& 0.85&0.45 &  9.57& 0.815\\
  & \textit{J=12} & 0.85 &0.48 &      0.86 &0.40& 0.80&0.41 &7.65&0.621\\
  &\textit{J=21}  & 0.90 &0.51 &      0.90 &0.38& 0.76&0.38 &6.89&0.410\\
   &\textit{J=30}   & 0.92 &0.54 &      0.92 &0.34& 0.70&0.32 &5.45&0.398\\
   & \textit{J=99}  & 0.95& 0.57 &    0.93 &0.15  & 0.61&0.16&3.55&0.254\\
   & \textit{J=252}  & 0.99& 0.58 &    0.99 &0.15  & 0.60&0.10&1.78&0.141\\
 \bottomrule 
	\end{tabular}
\vspace{-1mm}
\end{table}

\textit{Setting 2}.
More complex structures are considered with $K=5$. Two scenarios are considered,  where in the first scenario, all the factors are identifiable, while
in the second scenario, there are some factors that cannot be identifiable. To be specific, in scenario 1,
\begin{equation}
Q_1^{\top} =
\left( \begin{array}{cccccccccccc}
1 & 0 & 0 & 1 & 1 &\cdots &1 & 0 & 0 & 1 & 1 &\cdots\\
1 & 1 & 0 & 0 & 1 &\cdots &1 & 1 & 0 & 0 & 1 &\cdots\\
1 & 1 & 1 & 0 & 0 &\cdots &1 & 1 & 1 & 0 & 0 &\cdots\\
0 & 1 & 1 & 1 & 0 &\cdots &0 & 1 & 1 & 1 & 0 &\cdots\\
0 & 0 & 1 & 1 & 1 &\cdots &0 & 0 & 1 & 1 & 1 &\cdots\\
\end{array} \right),
\notag
\end{equation}
all latent factors are structurally identifiable with  with a mixed structure,
where $|R_{Q_1}(S)| = J/5, S = \{1, 2, 3\},\{2, 3, 4\},\{3, 4, 5\},\{4, 5, 1\}, \text{and} \{5, 1, 2\}.$ 
In scenario 2, 
\begin{equation}
Q_2^{\top} =
\left( \begin{array}{cccccccccccc}
1 & 0 & 0 & 1 & 1 &\cdots &1 & 0 & 0 & 1 & 1 &\cdots\\
1 & 1 & 0 & 0 & 1 &\cdots &1 & 1 & 0 & 0 & 1 &\cdots\\
1 & 1 & 1 & 0 & 0 &\cdots &1 & 1 & 1 & 0 & 0 &\cdots\\
\textbf{1} & 1 & 1 & 1 & 0 &\cdots &\textbf{1} & 1 & 1 & 1 & 0 &\cdots\\
0 & 0 & 1 & 1 & 1 &\cdots &0 & 0 & 1 & 1 & 1 &\cdots\\
\end{array} \right),
\notag
\end{equation}
only the third latent factor is not identifiable, the design structure is given by $|R_{Q_2}(S)| = J/5, S = \{1, 2, 3,4\},\{2, 3, 4\},\{3, 4, 5\},\{4, 5, 1\}, \text{and} \{5, 1, 2\}.$

\begin{table}[ht]
\renewcommand\arraystretch{0.9}
  \setlength{\tabcolsep}{0.8mm}
	\centering
	\begin{tabular}{@{}cccccccc@{}} 	\toprule 
	&   & ${\rm Corr}_{\ve x_1}$ & ${\rm Corr}_{\ve x_3}$& ${\rm Sin}_{\ve x_1}$ & ${\rm Sin}_{\ve x_3}$ & $d_{\ve X\ve A}$ & $d_f$ \\  
	\midrule
 \multirow{7}{*}{\textit{NSLFA}}
  & \textit{J=10}  &      0.70&0.71& 0.65&0.60 & 35.45& 0.915\\
  & \textit{J=50}  &      0.75 &0.78& 0.52&0.56 &27.10&0.821\\
  &\textit{J=100}  &      0.82 &0.83& 0.46&0.43 &14.21&0.610\\
   &\textit{J=150}   &      0.87 &0.90& 0.34&0.32 &13.65&0.498\\
   & \textit{J=200} &    0.93 &0.95  & 0.17&0.16&5.73&0.410\\
   & \textit{J=300}  &    0.99 &0.99 & 0.12&0.16&0.25&0.310\\
     & \textit{J=500}  &    0.99 &0.99  & 0.12&0.16&0.15&0.110\\

    \midrule
 \multirow{7}{*}{\textit{Linear FA}}
 & \textit{J=10}  &      0.51&0.50& 0.85&0.82 & 45.55& \\
  & \textit{J=50}  &      0.57 &0.60& 0.71&0.77 &39.10&\\
  &\textit{J=100}  &      0.66&0.67&0.69&0.65 &26.81&\\
   &\textit{J=150}   &     0.78 &0.80& 0.47&0.50&16.61&\\
   & \textit{J=200} &   0.80 &0.80 & 0.44&0.44&9.56&\\
   & \textit{J=300}  &    0.82 &0.81  & 0.41&0.41&8.87&\\
     & \textit{J=500}  &    0.82 &0.81  & 0.41&0.38&8.74&\\

     \midrule
 \multirow{3}{*}{\textit{GPLVM}}
 & \textit{J=10}  &      0.51&0.10& 0.85&0.99 & & 0.780\\
  & \textit{J=50}  &      0.57 &0.15& 0.81&0.99 &&0.610\\
  &\textit{J=100}  &      0.60&0.11&0.80&0.99 &&0.410\\

 \bottomrule 
	\end{tabular}
\vspace{-1mm}
\caption{The result of \textit{NSLFA}, \textit{Linear FA} and \textit{GPLVM} in Setting 2}
\label{Example3}
\end{table}

\begin{table}[ht]
\renewcommand\arraystretch{0.9}
  \setlength{\tabcolsep}{0.8mm}
	\centering
	\begin{tabular}{@{}cccccccc@{}} 	\toprule 
	 &   & ${\rm Corr}_{\ve x_1}$ & ${\rm Corr}_{\ve x_3}$& ${\rm Sin}_{\ve x_1}$ & ${\rm Sin}_{\ve x_3}$ & $d_{\ve X\ve A}$ & $d_f$ \\

 \midrule
 \multirow{7}{*}
  & \textit{J=10}  &      0.70&0.41& 0.65&0.85 & 30.82& 0.915\\
  & \textit{J=50}  &      0.75 &0.45& 0.52&0.81 &26.33&0.721\\
  &\textit{J=100}  &      0.82 &0.49& 0.46&0.78 &15.77&0.610\\
   &\textit{J=150}   &      0.87 &0.52& 0.34&0.61 &11.56&0.498\\
   & \textit{J=200} &    0.93 &0.55  & 0.17&0.54&4.33&0.378\\
   & \textit{J=300}  &    0.99 &0.57  & 0.12&0.52&2.12&0.260\\
     & \textit{J=500}  &    0.99 &0.58  & 0.12&0.51&1.71&0.110\\

 \bottomrule 
	\end{tabular}
\vspace{-1mm}
\caption{The result of  \textit{scenario 2} in Setting 2}
\label{Example3S2}
\end{table}

\section{Technical details}
\subsection{Regular conditions}
\label{A1A2}
We now characterize the structural identifiability under suitable regularity conditions. Recall that design matrix $Q \in\{0,1\}^{\mathbb{Z}\times\Bar{k}}$. Our first regularity assumption is about the stability of the $Q$ matrix.

\textbf{A1} The set
$R_Q(S)=\left\{j: q_{j k}=1, \text { if } k \in S \text { and } q_{j k}=0,\right. 
\text { if } k \notin S, 1 \leq j \leq J\}$ is non-empty for any subset $S \subset\{1, \ldots, K\}$. 

\textbf{A2} 
The parameter space $\mathcal{S}_Q \subset \mathbb{R}^{\mathbb{Z}_{+} \times
K} \times$ $\mathbb{R}^{J\times K}$,
$
\mathcal{S}_Q=\mathcal{S}_Q^{(1)} \times \mathcal{S}_Q^{(2)}
$,
here we define
$$
\mathcal{S}_Q^{(1)}=\left\{\ve X \in \mathbb{R}^{\mathbb{Z}_{+} \times K}:\left\|\boldsymbol{x}_i\right\| \leq C \text { and } \gamma(\ve X)>0\right\}
$$
and
$$
\begin{aligned}
\mathcal{S}_Q^{(2)}=\{ & \ve A \in \mathbb{R}^{J \times K}:\left\|\ve{a}_j\right\| \leq C,~~\text{the columns of }~~\ve A_{\left[R_Q(S), S\right]}~~\text {are linearly independent} \},
\end{aligned}
$$
where $C$ is a positive constant, $\ve A_{\left[R_Q(S), S\right]}$ is a submatrix of $\ve A$ consisting of the rows indexed by $R_Q(S)$ and the columns indexed by $S$, the $\gamma$ function is defined as:
 $$
\gamma(\ve X) \triangleq \liminf _{n \rightarrow \infty} \frac{\sigma_m\left(\ve X_{[1: n, 1: m]}\right)}{\sqrt{n}},
$$ 
and
$\sigma_1(\ve X) \geq \sigma_2(\ve X) \geq \cdots \geq \sigma_m(\ve X)$ are the singular values of a matrix $\ve X \in \mathbb{R}^{n \times m}$, in a descending order.

\textbf{A3} $\phi_k(\boldsymbol{a}_j^{\top} \ve x_i)$ defined in the proof of Theorem 2 in Appendix \ref{pf1} is thrice differentiable with respect to $\boldsymbol{a}_j^{\top} \ve x_i$ for each $j\in\{1,2,\ldots, J\}$.

\textbf{A4} $f_j^*$ is continuously differentiable on a compact set.

\subsection{Proofs}
\label{pf1}
{\bf Proof of Theorem 1:} Proof by contradiction.  If $
\{k\}=\bigcap_{k \in S, \ R_Q(S)\ \text{non-empty}} S,
$ is not satisfied, then
there are two cases: (1) $\left\{k, k^{\prime}\right\} \subset \bigcap_{k \in S, \ R_Q(S)\ \text{non-empty}} S$ for some $k^{\prime} \neq k$, and (2) $\emptyset =\bigcap_{k \in S, \ R_Q(S)\ \text{non-empty}} S$. For these two cases, we follow \cite{chen2020structured}
to construct $(\tilde{\boldsymbol{X}}, \tilde{\ve A}),\left(\boldsymbol{\ve X}^{\prime}, \ve A^{\prime}\right) \in \mathcal{S}_Q$.

According to the proof of Proposition 1 in \cite{chen2020structured}, we 
construct $(\ve X, \ve A) \in \mathcal{S}_Q$ where  $
\ve X=C\left[\begin{array}{c}
I_K \\
I_K \\
\vdots
\end{array}\right]
$. For each $S$, when $R_Q(S)$ is a empty set, we construct $\ve A_{\left[R_Q(S), 1: K\right]}=\ve{0}$. When $R_Q(S)$ is \ non-empty, we construct 
$
\ve A_{\left[R_Q(S), S\right]}=C\left[\begin{array}{c}
I_{|S|} \\
I_{|S|} \\
\vdots
\end{array}\right]
$, where $I_m$  denotes the $m\times m$ Identity matrix.

\textbf{Case 1}: $\left\{k, k^{\prime}\right\} \subset \bigcap_{k \in S, \ R_Q(S)\ \text{non-empty}} S$.  Let $\tilde{\ve X}=\frac{1}{2} \ve X$, $\tilde{\ve A}=\frac{1}{2} \ve A$, and the $\left(\ve X^{\prime}, \ve A^{\prime}\right)$ are constructed  as follows,
$$
x_{i m}^{\prime}=\left\{\begin{array}{l}
x_{i m} / 2 \text { if } m \neq k \\
\left(x_{i k}-x_{i k^{\prime}}\right) / 2 \text { if } m=k
\end{array} \quad \text { and } a_{j m}^{\prime}=\left\{\begin{array}{l}
a_{j m} / 2 \text { if } m \neq k^{\prime} \\
\left(a_{j k}+a_{j k^{\prime}}\right) / 2 \text { if } m=k^{\prime},
\end{array}\right.\right.
$$
It is easy to check that  $\sum_{m=1}^K x_{i m}^{\prime} a_{j m}^{\prime}=\sum_{m=1}^K \tilde{x}_{i m} \tilde{a}_{j m}$ for all $i\in \mathbb{Z}_{+},\ j\in\{1,\ldots, J\}$, i.e., $\tilde{\boldsymbol{a}}_j^{\top} \tilde{\ve x}_i={\boldsymbol{a}_j^{\top}}^{\prime}\boldsymbol{x}_i^{\prime}$ for all $i\in \mathbb{Z}_{+},\ j\in\{1,\ldots, J\}$. 
Since $
\ve Y_{j} \mid \ve a_j, \ve x_i, \tau, w, \sigma^2 \sim N\left(\ve{0}_N, \sigma^2 \boldsymbol{I}_N+\boldsymbol{C}_{j}\right)
$
where $$\boldsymbol{C}_{j}(i, l)=\operatorname{Cov}\left(f\left(\ve a_j^{\top} \ve x_i\right), f\left(\ve a_j^{\top} \ve x_l\right)\right)=\tau \exp \left(-\frac{w\left(\boldsymbol{a}_j^{\top} \ve x_i-\boldsymbol{a}_j^{\top} \ve x_l\right)^2}{2}\right),$$
when given the hyper-parameters $\tau, w$, this can  lead to $P_{\left(\tilde{\ve X}, \tilde{\ve A}\right)}=P_{\left(\ve X^{\prime}, \ve A^{\prime}\right)}$, where $P_{\left(\ve X, \ve A\right)}$ denote the  probability distribution of $\left(Y_{i j}, i\in \mathbb{Z}_{+},\ j\in\{1,\ldots, J\}\right)$ given the parameters $ \ve X,  \ve A,  \tau, w, \sigma^2$.
From the proof of Theorem 1 in \cite{chen2020structured}, we can see that 
$\sin _{+} \angle\left(\tilde{\ve X}_{[k]}, \ve X_{[k]}^{\prime}\right)=1 / \sqrt{2}>0$. This contradicts the definition of structural identifiability.

\textbf{Case 2}: We assume $K \geqslant 2$.  Let $\tilde{\ve X}, \tilde{\ve A}$ be constructed in the same way as we did in case 1. Further let $\ve A^{\prime}=\tilde{\ve A}$,   $\ve X^{\prime}$ is constructed as follows: 
$\ve X_{[m]}^{\prime}=\frac{1}{2} \tilde{\ve X}_{[m]}$ for $m \geqslant 2$, and $x_{i 1}^{\prime}=C / 2$ for all $i \in \mathbb{Z}_{+}$. 
It is easy to check that $\ve{a}_j^{\prime \top} \boldsymbol{x}_i^{\prime}=\tilde{\ve{a}}_j^{\top} \tilde{\boldsymbol{x}}_i$ for all $i\in \mathbb{Z}_{+},\ j\in\{1,\ldots, J\}$ since $
\ve Y_{j} \mid \ve a_j, \ve x_i, \tau, w, \sigma^2 \sim N\left(\ve{0}_N, \sigma^2 \boldsymbol{I}_N+\boldsymbol{C}_{j}\right)
$, then $P_{(\tilde{\ve X}, \tilde{\ve A})}=P_{(\ve X^{\prime}, \ve A^{\prime})}$. From the proof of Theorem 1 in \cite{chen2020structured}, we have $\sin _{+} \angle\left(\tilde{\ve X}_{[1]}, \ve X_{[1]}^{\prime}\right)=\sqrt{K-1/K}>0$. By contradiction, the proof of Theorem 1 completes.

\textbf{Proof of \eqref{Thm2.1} in Theorem 2} 
Based on Theorem 2.1 of \cite{Basawa:1980}, let $\theta=\boldsymbol{a}_j^{\top} \ve x_i$, then we find the density function $p_k(\theta)$. For fixed $j$, assuming other parameters are given, the marginal likelihood of $\boldsymbol{a}_j^{\top} \ve x=(\boldsymbol{a}_j^{\top} \ve x_1,\ldots,\boldsymbol{a}_j^{\top} \ve x_N)$, i.e., the marginal distribution of $\ve Y_{j}=$ $\left(Y_{1j}, \ldots, Y_{Nj}\right)^{\top}, N \geq 1$ in the GP-SIM factor analysis model in Equation \ref{GPSIMFA},  has a multivariate normal distribution with mean $\ve{0}$ and covariance $\sigma^2 \boldsymbol{I}_N+\boldsymbol{C}_{j}$ as in Equation \ref{distributionYn}. Note that $\ve{y}^k$ has nonsingular normal distribution $N_k\left(\ve{0}_m, \boldsymbol{\Sigma}_k\right)$. Thus, applying the standard theory of multivariate normal distribution, $p_k(\boldsymbol{a}_j^{\top} \ve x_i)$, the conditional probability density of $Y_k$ given $\ve Y^{k-1}$ is also a normal density with mean $m_k(\boldsymbol{a}_j^{\top} \ve x)$ and variance $v_k(\boldsymbol{a}_j^{\top} \ve x)$, where $m_k(\boldsymbol{a}_j^{\top} \ve x)$ and $v_k(\boldsymbol{a}_j^{\top} \ve x)$ are some functions of $\theta$, determined by the linear combination of the submatrices of $\boldsymbol{\Sigma}_k$ and its inverse.
Then by calculation, $\phi_k(\boldsymbol{a}_j^{\top} \ve x_i)$ and its derivatives are given by
$$
\begin{aligned}
&\phi_k(\boldsymbol{a}_j^{\top} \ve x_i)
=\log p_k(\boldsymbol{a}_j^{\top} \ve x_i)
=-\log \left(\sqrt{2 \pi v_k(\boldsymbol{a}_j^{\top} \ve x_i)}\right)-\left\{\frac{1}{2 v_k(\boldsymbol{a}_j^{\top} \ve x_i)}\left(y_k-m_k(\boldsymbol{a}_j^{\top} \ve x_i)\right)^2\right\}, \\
 &\dot{\phi}_k(\boldsymbol{a}_j^{\top} \ve x_i)
=\frac{d}{d \boldsymbol{a}_j^{\top} \ve x_i} \phi_k(\boldsymbol{a}_j^{\top} \ve x_i)
 =-\frac{v_k^{\prime}(\boldsymbol{a}_j^{\top} \ve x_i)}{2v_k(\boldsymbol{a}_j^{\top} \ve x_i)}+\frac{v_k^{\prime}(\boldsymbol{a}_j^{\top} \ve x_i)}{2 v_k(\boldsymbol{a}_j^{\top} \ve x_i)^2}\left(y_k-m_k(\boldsymbol{a}_j^{\top} \ve x_i)\right)^2 \\
&\quad\quad\quad\quad\quad\quad\quad\quad\quad\quad\quad\quad\quad\ \ +\frac{\left(y_k-m_k(\boldsymbol{a}_j^{\top} \ve x_i)\right)}{v_k(\boldsymbol{a}_j^{\top} \ve x_i)} m_k^{\prime}(\boldsymbol{a}_j^{\top} \ve x_i), \\
&\ddot{\phi}_k(\boldsymbol{a}_j^{\top} \ve x_i)
 =\frac{d}{d \boldsymbol{a}_j^{\top} \ve x_i} \dot{\phi}_k(\boldsymbol{a}_j^{\top} \ve x_i)
 =A_k(\boldsymbol{a}_j^{\top} \ve x_i)\left(y_k-m_k(\boldsymbol{a}_j^{\top} \ve x_i)\right)^2\\
&\quad\quad\quad\quad\quad\quad\quad\quad\quad\quad\quad\quad\quad\ \ +B_k(\boldsymbol{a}_j^{\top} \ve x_i)\left(y_k-m_k(\boldsymbol{a}_j^{\top} \ve x_i)\right)+C_k(\boldsymbol{a}_j^{\top} \ve x_i),
\end{aligned}
$$
where $A_k(\boldsymbol{a}_j^{\top} \ve x_i), B_k(\boldsymbol{a}_j^{\top} \ve x_i)$, and $C_k(\boldsymbol{a}_j^{\top} \ve x_i)$ are some functions of $\boldsymbol{a}_j^{\top} \ve x_i$, made up of first and second derivatives of $m_k(\boldsymbol{a}_j^{\top} \ve x_i)$ and $v_k(\boldsymbol{a}_j^{\top} \ve x_i)$.
Subsequently, we apply a similar methodology as used in \cite{choi2011gaussian} to verify the conditions C1-C4 in \cite{Basawa:1980}, thus establishing the consistency of $\widehat{\ve a}_j^{\top}\widehat{ \ve x_i}$.

Applying a identical verification steps same as $\boldsymbol{a}_j^{\top} \ve x_i$ to other hyperparameters, $\tau$ and $w$ yields the consistency of the respective parameters. This completes the proof of \eqref{Thm2.1} in Theorem 2.

\textbf{Proof of \eqref{Thm2.2} in Theorem 2.} 
Recall the definition of $\mathcal{S}_Q$,  there exist $N_0$ and $\epsilon'>0$ such that for $N \geqslant N_0$,
$\sigma_K\left(\ve X^*\right) \geqslant \epsilon' \sqrt{N} $. For fixed $J$, 
$ \sigma_{|S|}\left(\ve A_{\left[R_Q(S) , S\right]}^*\right) \geqslant \epsilon' \sqrt{J}.$ From the structural identifiability condition of the $k$-th latent fatcor (Theorem 1), for all $S \subset\{1, \ldots, K\}$ such that $\ R_Q(S)\ \text{non-empty}$. 
Let  $\widehat{\ve M}=\widehat{\ve X} \widehat{\ve A}^{\top} \text { and } \ve M^*=\ve X^* \ve A^{* \top}$ and based on above results, we 
define an event
$$
\mathcal{E}=\left\{\left\|\widehat{\ve M}-\ve M^*\right\|_2<\frac{\sqrt{N J} \epsilon'^2}{2}\right\} .
$$
Then, using the fact that  $\sin \angle\left(\ve X{[k]}^*, \widehat{\ve X}_{[k]}\right) \leqslant 1,$ 

\begin{equation}
    \E \sin \angle\left(\ve X{[k]}^*, \widehat{\ve X}_{[k]}\right) \leqslant \E\left[\sin \angle\left(\ve X{[k]}^*, \widehat{\ve X}_{[k]}\right) \ve{1}_{\mathcal{E}}\right]+\Pr\left(\mathcal{E}^c\right). 
    \label{EsinEpsilon}
\end{equation}
From \eqref{Thm2.1} in Theorem 2, we have $\widehat{\ve a}_j^{\top}\widehat{ \ve x_i} \stackrel{p}{\rightarrow} \boldsymbol{a}_{j*}^{\top} \ve x_{i*} \ \text{for} \ j=1,\ldots,J, \text{as}\ N{\rightarrow}\infty$, which indicates that  
\begin{equation}
    \Pr\left(\mathcal{E}^c\right) = \Pr\left(\left\|\widehat{\ve M}-\ve M^*\right\|_2\geq \frac{\sqrt{N J} \epsilon'^2}{2}\right) \leqslant \frac{2E\left\|\widehat{\ve M}-\ve M^*\right\|_2}{\sqrt{N J}\epsilon'^2} \leqslant \frac{2E\left\|\widehat{\ve M}-\ve M^*\right\|_F}{\sqrt{N J}\epsilon'^2} {\rightarrow} 0.
    \label{MFroupper}
\end{equation}

We proceed to the analysis of $\E\left[\sin \angle\left(\ve X_{[k]}^*, \widehat{\ve X}_{[k]}\right) \ve{1}_{\mathcal{E}}\right]$. Recall that when the event $\mathcal{E}$ happens, we have
\begin{equation}
    2\left\|\widehat{\ve M}-\ve M^*\right\|_2 <\sqrt{N J} \epsilon'^2,
    \label{pfthm2Episilon}
\end{equation}
Based on \eqref{EsinEpsilon},\eqref{MFroupper} \eqref{pfthm2Episilon} and the
proof of Theorem 2 in \citep{chen2020structured}, we can get
\begin{equation}
    \E\sin \angle\left(\ve X_{[k]}^*, \widehat{\ve X}_{[ k]}\right) \leqslant C_1 \frac{\left\|\widehat{\ve M}-\ve M^*\right\|_F}{\sqrt{N J}} {\rightarrow} 0\ \text{as}\ N{\rightarrow}\infty.\notag
\end{equation}
This completes the proof of \eqref{Thm2.2} in Theorem 2.

\textbf{Proof of Theorem 3.} From Theorem 2, we have $
\widehat{\ve a}_j^{\top}\widehat{ \ve x_i} \stackrel{p}{\rightarrow} \boldsymbol{a}_{j*}^{\top} \ve x_{i*}, \widehat{\tau} \stackrel{p}{\rightarrow} \tau_*,   \widehat{w} \stackrel{p}{\rightarrow} w_*.
$ The following two consistent estimator $\widehat{\tau}, \widehat{w}$ are treated as fixed hyperparameters in the covariance function. And since $f_{j*}$ is continuous, it follows that $f_{j*}\left(\widehat{\ve a}_j^{\top}\widehat{ \ve x_i}\right) \stackrel{p}{\rightarrow} f_{j*}\left(\boldsymbol{a}_j^{\top} \ve x_i\right).$ Using triangle inequality, 
\begin{equation}\left| f_j(\widehat{\ve a}_j^{\top}\widehat{ \ve x_i}) - f_{j*}(\boldsymbol{a}_{j*}^{\top} \ve x_{i*}) \right| <\left| f_j(\widehat{\ve a}_j^{\top}\widehat{ \ve x_i}) - f_{j*}(\widehat{\ve a}_j^{\top}\widehat{ \ve x_i}) \right|+ \left| f_{j*}(\widehat{\ve a}_j^{\top}\widehat{ \ve x_i}) - f_{j*}(\boldsymbol{a}_{j*}^{\top} \ve x_{i*}) \right|.  \label{triangel}
\end{equation}
According to the above discussion, the second term on the right converges to 0 in probability. To make the conclusion hold, we only need to prove that for every $\epsilon>0$, 
$$
\Pi\left\{(f_j, \sigma) \in W_\epsilon^c \mid \boldsymbol{Y},  \widehat{\ve a}_j^{\top}\widehat{ \ve x_i},  \widehat{\tau}, \widehat{w}\right\} \rightarrow 0, \text { a.s.} [P_{f_{j*},\sigma_*}],
$$
where
$$
W_\epsilon=\left\{(f_j, \sigma): \left|f_j\left(\widehat{\ve a}_j^{\top}\widehat{ \ve x_i}\right)-f_{j*}\left(\widehat{\ve a}_j^{\top}\widehat{ \ve x_i}\right)\right| <\epsilon,\left|\frac{\sigma}{\sigma_*}-1\right|<\epsilon\right\},
$$
$ W_\epsilon^c$ is the complementary set of $ W_\epsilon.$

The above result follows from Theorem 1 in \cite{choi2007posterior} after setting $\theta=(f_j, \sigma), \theta_*=(f_{j*}, \sigma_*), Z_i=Y_i \mid f_j, \sigma^2, \widehat{\ve a}^{\top}_j\widehat{\ve x_i}$, and verifying their conditions (A1) and (A2). By calculation,   
$$
K_i\left(\theta_* ; \theta\right)=\frac{1}{2} \log \frac{\sigma^2}{\sigma_*^2}-\frac{1}{2}\left(1-\frac{\sigma_*^2}{\sigma^2}\right)+\frac{1}{2} \frac{\left[f_{j*}\left(\widehat{\ve a}_j^{\top}\widehat{ \ve x_i}\right)-f_j\left(\widehat{\ve a}_j^{\top}\widehat{ \ve x_i}\right)\right]^2}{\sigma^2}
$$
and
$$
V_i\left(\theta_*, \theta\right)=2\left[-\frac{1}{2}+\frac{1}{2} \frac{\sigma_*^2}{\sigma^2}\right]^2+\left[\frac{\sigma_*^2}{\sigma^2}\left[f_j\left(\widehat{\ve a}_j^{\top}\widehat{ \ve x_i}\right)-f_{j*}\left(\widehat{\ve a}_j^{\top}\widehat{ \ve x_i}\right)\right]\right]^2
$$
For each $\delta>0$, define
$$
B_\delta=\left\{(f_j, \sigma):\left\|f_j-f_{j*}\right\|_{\infty}<\delta,\left|\frac{\sigma}{\sigma_*}-1\right|<\delta\right\} .
$$
Then, from the calculations of $K_i\left(\theta_*, \theta\right)$ and $V_i\left(\theta_*, \theta\right)$, it is easy to verify that condition (A1) holds. The condition (A2) is to verify the existence of test. Based on Assumption A4 and the model structure, following \cite{choi2007posterior} and \cite{choi2011gaussian}, the test is constructed. This  completes the proof of Theorem 3.

\end{document}